\documentclass[aps,prd,amsmath
,eqsecnum 
,preprintnumbers
,nofootinbib 
 ,showpacs 
 ,showkeys 
,twocolumn 
]{revtex4}
\usepackage[dvips]{graphicx}
\usepackage{amsmath}
\usepackage{amsfonts}
\usepackage{amssymb}
\usepackage{longtable} 
\usepackage{color}

\allowdisplaybreaks[4] 

\newcommand{\Chr}[3]{\mbox{\small\( \begin{Bmatrix}#1\\#2#3\end{Bmatrix}\)}}

\newcommand{\df}{\ {\overset {\rm def} =}\ }
\newcommand{\dr}[2]{\frac {{\rm d} {#1}} {{\rm d} {#2}}}

\newcommand{\dril}[2]{{{\rm d} {#1}} / {{\rm d} {#2}}}

\begin{document}

\title{Short-lived flashes of gamma radiation in a quasi-spherical Szekeres metric}

\author{Andrzej Krasi\'nski}
\affiliation{N. Copernicus Astronomical Centre, Polish Academy of Sciences, \\
Bartycka 18, 00 716 Warszawa, Poland} \email{akr@camk.edu.pl}

\date {}

\begin{abstract}
In previous papers it was shown that gamma rays with characteristics similar to
those of the gamma-ray bursts (GRBs) observed by astronomers may arise from
suitably shaped nonuniformities in the Big Bang in a quasi-spherical Szekeres
(QSS) model. The gamma radiation arises by blueshifting the light emitted by
hydrogen atoms at the end of the last scattering epoch along preferred
directions that exist in QSS models. However, the durations of the gamma flashes
and of their afterglows implied by the model were much longer than those of the
observed GRBs. In this paper it is shown that for the gamma-ray flash a duration
of correct order results if the blueshifted radiation, on its way to the present
observer, passes through another QSS region where it is deflected. The angle of
deflection changes with time because of the cosmological drift mechanism, so the
high-frequency ray will miss the observer after a while. It is shown by explicit
numerical calculation that a gamma-ray flash will no longer be visible to the
present observer after 10 minutes.
\end{abstract}

\maketitle

\section{Motivation and background}\label{intro}

\setcounter{equation}{0}

In previous papers by this author \cite{Kras2016a,Kras2016b,Kras2017} it was
shown that flashes of gamma radiation with characteristics similar to those of
the gamma-ray bursts (GRBs), now routinely observed by astronomers
\cite{Perlwww}, may arise from a non-simultaneous Big Bang (BB) in a
quasi-spherical Szekeres (QSS) model \cite{Szek1975,Szek1975b,PlKr2006}. The
gamma radiation arises by blueshifting the light emitted by hydrogen atoms at
the end of the last scattering epoch along preferred directions existing in QSS
models \cite{Kras2016b,Kras2017} (for the first discussions of blueshifting see
Refs. \cite{Szek1980,HeLa1984}). With the BB profile $t_B(r)$ chosen suitably,
the blueshift is sufficiently strong to move the observed frequency of the
radiation to the gamma range. A multitude of sources can be created by
distributing many small QSS regions over a Friedmann background \cite{Kras2017}.
However, the durations of the gamma flashes and of their afterglows implied by
the model were much longer than those measured for the GRBs.

In this paper it is shown that a duration of the gamma-ray flash of correct
order results if the blueshifted radiation, on its way to the present observer,
passes through another QSS region, where it is deflected. The angle of
deflection changes with time in consequence of the cosmic drift
\cite{KrBo2011,QABC2012,KoKo2017}, so the high-frequency ray will miss the
observer after a while.

Sections \ref{QSSS} -- \ref{ERS} are partly repeated after Ref. \cite{Kras2017};
they present the QSS model used in this paper (Sec. \ref{QSSS}), the null
geodesic equations and properties of redshift along them (Sec. \ref{symmetric})
and the definition of the extremum redshift surface (ERS, Sec. \ref{ERS}).

In Sec. \ref{deflect}, the configuration of the QSS regions is presented, and
the time-dependent deflection of light rays in them is demonstrated on a
numerical example. In Sec. \ref{drift} the rate of angular drift of the
deflected rays is calculated. In Sec. \ref{durGRB} it is shown by explicit
numerical calculation that in the configuration of Sec. \ref{deflect} the
present observer who registered a gamma-radiation flash at time $t_o$ will no
longer see it at $t'_o = t_o +$ 10 minutes. Instead, the radiation coming from
the same direction at $t'_o$ will have its frequency in the ultraviolet range.
(The 10 minutes were chosen as an exemplary duration of an observed GRB
\cite{Perlwww}, but the gamma-ray flash is, in this model, instantaneous; see
Sec. \ref{sumup}.)

In Sec. \ref{nearby} it is shown that the brief duration of the gamma-ray flash
is preserved if the initial point and direction of the later-arriving ray are
slightly perturbed. Thus, the result of Sec. \ref{durGRB} is not overly
sensitive to numerical inaccuracies. Section \ref{sumup} is a summary of the
results and of the problems that remain to be solved. The appendices present
some details of the computations.

\section{The quasispherical Szekeres (QSS) spacetime used in this
paper}\label{QSSS}

\setcounter{equation}{0}

The signature and labelling of coordinates will be $(+, -, -, -)$ and
$\left(x^0, x^1, x^2, x^3\right) = (t, r, x, y)$ or $(t, r, \vartheta,
\varphi)$.

The metric of the QSS spacetimes is \cite{Szek1975,Szek1975b,PlKr2006,Hell1996}
\begin{equation}\label{2.1}
{\rm d} s^2 = {\rm d} t^2 - \frac {\left(\Phi,_r - \Phi {\cal E},_r/{\cal
E}\right)^2} {1 + 2 E(r)} {\rm d} r^2 - \left(\frac {\Phi} {\cal E}\right)^2
\left({\rm d} x^2 + {\rm d} y^2\right),\ \ \ \ \
\end{equation}
where
\begin{equation}\label{2.2}
{\cal E} \df \frac S 2 \left[\left(\frac {x - P} S\right)^2 + \left(\frac {y -
Q} S\right)^2 + 1\right],
\end{equation}
$P(r)$, $Q(r)$, $S(r)$ and $E(r)$ being arbitrary functions such that $S \neq 0$
and $E \geq -1/2$ at all $r$.

The source in the Einstein equations is dust ($p = 0$) with the velocity field
$u^{\alpha} = {\delta_0}^{\alpha}$. The surfaces of constant $t$ and $r$ are
nonconcentric spheres, and $(x, y)$ are the stereographic coordinates on each
sphere. At a fixed $r$, they are related to the spherical coordinates by
\begin{eqnarray}\label{2.3}
x &=& P + S \cot(\vartheta/2) \cos \varphi, \nonumber \\
y &=& Q + S \cot(\vartheta/2) \sin \varphi.
\end{eqnarray}
The functions $(P, Q, S)$ determine the centres of the spheres in the spaces of
constant $t$ (see illustrations in Ref. \cite{Kras2016b}). Because of the
non-concentricity, the QSS spacetimes in general have no symmetry
\cite{BoST1977}.

With $\Lambda = 0$ assumed, $\Phi(t,r)$ obeys
\begin{equation}\label{2.4}
{\Phi,_t}^2 = 2 E(r) + \frac {2 M(r)} {\Phi},
\end{equation}
where $M(r)$ is an arbitrary function. We will consider only models with $E >
0$, then the solution of (\ref{2.4}) is
\begin{eqnarray}\label{2.5}
\Phi(t,r) &=& \frac M {2E} (\cosh \eta - 1), \nonumber \\
\sinh \eta - \eta &=& \frac {(2E)^{3/2}} M \left[t - t_B(r)\right],
\end{eqnarray}
where $t_B(r)$ is an arbitrary function; $t = t_B(r)$ is the time of the BB
singularity, at which $\Phi(t_B, r) = 0$. We assume $\Phi,_t > 0$ (the Universe
is expanding).

The mass density implied by (\ref{2.1}) is
\begin{equation}\label{2.6}
\kappa \rho = \frac {2 \left(M,_r - 3 M {\cal E},_r / {\cal E}\right)} {\Phi^2
\left(\Phi,_r - \Phi {\cal E},_r / {\cal E}\right)}, \quad \kappa \df \frac {8
\pi G} {c^2}.
\end{equation}
This is a mass-dipole superposed on a spherical monopole \cite{DeSo1985},
\cite{Szek1975b}. The dipole vanishes where ${\cal E},_r = 0$. The density is
minimum where ${\cal E},_r/{\cal E}$ is maximum and vice versa \cite{HeKr2002}.

The arbitrary functions must be such that no singularities exist after the BB.
This is ensured by \cite{HeKr2002}:
\begin{eqnarray}
\frac {M,_r} {3M} &\geq& \frac {\cal P} S, \qquad \frac {E,_r} {2E} > \frac
{\cal P} S ~~~~\forall~r, \label{2.7} \\
{\rm where}\ \ {\cal P} &\df& {\sqrt{(S,_r)^2 + (P,_r)^2 + (Q,_r)^2}}.
\label{2.8}
\end{eqnarray}
These inequalities imply \cite{HeKr2002}
\begin{equation}\label{2.9}
\frac {M,_r} {3M} \geq \frac {{\cal E},_r} {\cal E}, \qquad \frac {E,_r} {2E} >
\frac {{\cal E},_r} {\cal E} \qquad \forall~r.
\end{equation}

The extrema of ${\cal E},_r/{\cal E}$ with respect to $(x, y)$ are
\cite{HeKr2002}
\begin{equation}\label{2.10}
\left.\frac {{\cal E},_r} {\cal E}\right|_{\rm ex} = \varepsilon_2 \frac {\cal
P} S, \qquad \varepsilon_2 = \pm 1,
\end{equation}
with $+$ at a maximum and $-$ at minimum; they occur at
\begin{equation}\label{2.11}
x = P + \frac {SP,_r} {{\cal P} + \varepsilon_2 S}, \qquad y = Q + \frac {SQ,_r}
{{\cal P} + \varepsilon_2 S}.
\end{equation}

The Lema\^{\i}tre \cite{Lema1933} -- Tolman \cite{Tolm1934} (L--T) models are
contained in (\ref{2.1}) -- (\ref{2.2}) as the limit of constant $(P, Q, S)$.
The Friedmann limit is obtained from QSS when $E / M^{2/3}$ and $t_B$ are
constant (then $(P, Q, S)$ can be made constant by a coordinate transformation).
QSS and Friedmann spacetimes can be matched at any constant $r$.

Because of $p = 0$, the QSS models can describe the evolution of the Universe no
further back in time than to the last scattering hypersurface (LSH); see Sec.
\ref{deflect}.

We will consider such QSS spacetimes whose L--T limit is Model 2 of Ref.
\cite{Kras2016a}. The $r$-coordinate is chosen so that
\begin{equation}\label{2.12}
M = M_0 r^3,
\end{equation}
and $M_0 = 1$ (kept in formulae for dimensional clarity) \cite{Kras2014d}. The
function $E(r)$, assumed in the form
\begin{equation}\label{2.13}
2E/r^2 \df -k = 0.4,
\end{equation}
is the same as in the background Friedmann model.

The units used in numerical calculations were introduced and justified in Ref.
\cite{Kras2014a}. Taking \cite{unitconver}
\begin{equation}\label{2.14}
1\ {\rm pc} = 3.086 \times 10^{13}\ {\rm km}, \quad 1\ {\rm y} = 3.156 \times
10^7\ {\rm s},
\end{equation}
the numerical length unit (NLU) and the numerical time unit (NTU) are defined as
follows:
\begin{equation}\label{2.15}
1\ {\rm NTU} = 1\ {\rm NLU} = 9.8 \times 10^{10}\ {\rm y} = 3 \times 10^4\ {\rm
Mpc}.
\end{equation}

\begin{figure}[h]
\includegraphics[scale=0.5]{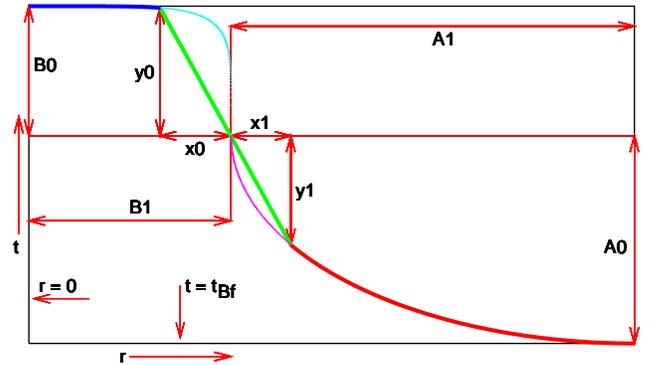}
\caption{Parameters of the bang-time profile in the quasi-spherical Szekeres
region; see text for explanation.} \label{drawpicture6}
\end{figure}

The BB profile is the same as in Ref. \cite{Kras2016a}, see Fig.
\ref{drawpicture6}. It consists of two curved arcs and of a straight line
segment joining them. The upper-left arc is a segment of the curve
\begin{equation}\label{2.16}
\frac {r^6} {{B_1}^6} + \frac {\left(t - t_{\rm Bf} - A_0\right)^6} {{B_0}^6} =
1,
\end{equation}
where
\begin{equation}\label{2.17}
t_{\rm Bf} = -0.13945554689046649\ {\rm NTU} \approx -13.67 \times 10^9\ {\rm
years}.
\end{equation}
The $t_{\rm Bf}$ is the asymptotic value of $t_B(r)$ in the L--T model that
mimicked accelerating expansion \cite{Kras2014d,Kras2014a}. This differs by
$\sim 1.6 \%$ from $(- T)$, where $T$ is the age of the Universe determined by
the Planck satellite \cite{Plan2014}
\begin{equation}\label{2.18}
T = 13.819 \times 10^9\ {\rm y} = 0.141\ {\rm NTU}.
\end{equation}
The lower-right arc is a segment of the ellipse\footnote{In further figures it
looks like a vertical straight line segment because $A_1$ is extremely small.}
\begin{equation}\label{2.19}
\frac {\left(r - B_1 - A_1\right)^2} {{A_1}^2} + \frac {\left(t - t_{\rm Bf} -
A_0\right)^2} {{A_0}^2} = 1.
\end{equation}
The straight segment passes through the point where the full curves would meet;
its slope is determined by $x_0$.

\begin{figure}[h]
 \hspace{-4mm}
 \includegraphics[scale=0.5]{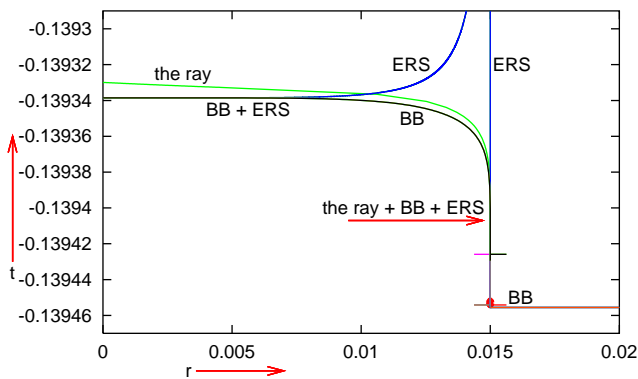}
\caption{The real BB profile. The horizontal strokes are at the ends of the
straight segment. The lower arc from Fig. \ref{drawpicture6} looks like a short
vertical straight line. ``The ray'' is one with the smallest $1 + z$; it
intersects the last scattering hypersurface at the dot. ERS is the extremum
redshift surface, see Sec. \ref{ERS}.}
 \label{realhump}
\end{figure}

The free parameters are $A_0$, $A_1$, $B_0$, $B_1$ and $x_0$. In Fig.
\ref{drawpicture6} the values of $x_0$ and $A_1$ are greatly exaggerated to
improve readability. The actual values are \cite{Kras2017}
\begin{equation} \label{2.20}
\left(\begin{array}{l}
A_0 \\
B_0 \\
A_1 \\
B_1 \\
x_0\\
\end{array}\right) = \left(\begin{array}{l}
 0.000026\ {\rm NTU} \\
 0.000091\ {\rm NTU} \\
 1 \times 10^{-10} \\
 0.015 \\
 10^{-11} \\
 \end{array}\right),
\end{equation}
and this BB profile is shown\footnote{Figure \ref{realhump} here is different
from Fig. 2 in Ref. \cite{Kras2017} -- the latter showed an intermediate profile
before final optimization.} in Fig. \ref{realhump}.

The QSS model is axially symmetric, with $P(r) = Q(r) = 0$ and
\begin{equation}\label{2.21}
S(r) = \sqrt{a^2 + r^2}
\end{equation}
(the same as in Ref. \cite{Kras2017}), where $a^2 = 0.001$. This $S(r)$ obeys
(\ref{2.7}), which, using (\ref{2.12}) -- (\ref{2.13}), reduces to
\begin{equation}\label{2.22}
1/r > S,_r / S.
\end{equation}
The axis of symmetry is at $x = y = 0$. At $r > r_F$, where
\begin{equation}\label{2.23}
r_F = A_1 + B_1 = 0.0150000001,
\end{equation}
the BB profile becomes horizontal straight, and the geometry of the model
becomes Friedmannian. See Sec. \ref{symmetric} for remarks on the choice of
coordinates in that region.

The Friedmann background is defined by
\begin{equation}\label{2.24}
{\rm d} s^2 = {\rm d} t^2 - {\mathcal R}^2(t) \left[\frac {{\rm d} r^2} {1 - k
r^2} + r^2 \left({\rm d} \vartheta^2 + \sin^2 \vartheta {\rm d}
\varphi^2\right)\right],\ \ \ \ \
\end{equation}
where $k$ and $t_{\rm Bf}$ are given by (\ref{2.13}) and (\ref{2.17}), and
\begin{equation}\label{2.25}
\Lambda = 0
\end{equation}
is assumed; $t = 0$ is the present time, $t_{\rm Bf}$ is the BB time. The scale
factor ${\mathcal R}(t)$ is determined by (\ref{2.5}) with $\Phi = r {\mathcal
R}(t)$, (\ref{2.12}) -- (\ref{2.13}) and $t_B(r) = t_{\rm Bf} =$ constant.

The mass-density at the last scattering time in the now-standard $\Lambda$CDM
model is \cite{Kras2016a}
\begin{equation}\label{2.26}
\kappa \rho_{\rm LS} = 56.1294161975316 \times 10^9 \ ({\rm NLU})^{-2}.
\end{equation}
In the L--T and Szekeres models it is assumed that a light ray intersects the
LSH when the density calculated along it becomes equal to the $\rho_{\rm LS}$ in
(\ref{2.26}).

With (\ref{2.24}), (\ref{2.25}), (\ref{2.13}) and (\ref{2.17}), the $\rho_{\rm
LS}$ of (\ref{2.26}) occurs at the redshift relative to the present time
\begin{equation}\label{2.27}
1 + z^b_{\rm LS} = 952.611615159.
\end{equation}
This differs by $\sim 12.7 \%$ from the $\Lambda$CDM value $z_{\rm LS} = 1090$
\cite{Plan2014,Plan2014b}. To bring our model to agreement with this, laborious
re-calculations would be required. Since the model needs improvements anyway, we
will rather stick to (\ref{2.25}), (\ref{2.13}) and (\ref{2.17}), to be able to
compare the present results with the earlier ones.

\section{Null geodesics in the axially symmetric QSS
spacetimes}\label{symmetric}

\setcounter{equation}{0}

In (\ref{2.1}) -- (\ref{2.2}) $x = \infty$ and $y = \infty$ are at the pole of
the stereographic projection of a sphere. This is a coordinate singularity where
numerical integration of geodesics breaks down. So, we introduce the coordinates
$(\vartheta, \varphi)$ by
\begin{equation}\label{3.1}
x = S_F \cot(\vartheta/2) \cos \varphi, \qquad y = S_F \cot(\vartheta/2) \sin
\varphi,
\end{equation}
where
\begin{equation}\label{3.2}
S_F \df S(r_F) = \sqrt{a^2 + {r_F}^2}
\end{equation}
is the value of $S$ at the Szekeres/Friedmann boundary. This changes (\ref{2.1})
and (\ref{2.2}) to
\begin{eqnarray}
{\rm d} s^2 &=& {\rm d} t^2 - \frac {{\cal N}^2 {\rm d} r^2} {1 + 2 E(r)} -
\left(\frac {\Phi} {\cal F}\right)^2 \left({\rm d} \vartheta^2 + \sin^2
\vartheta {\rm d} \varphi^2\right), \nonumber \\
&& \ \ \ \ \label{3.3}\\
{\cal F} &=& \frac {S_F} {2 S}\ (1 + \cos \vartheta) + \frac S {2 S_F}\ (1 -
\cos \vartheta), \label{3.4}
\end{eqnarray}
where

\begin{equation}\label{3.5}
{\cal N} \df \Phi,_r - \Phi {\cal F},_r/{\cal F}.
\end{equation}
In general, $(\vartheta, \varphi)$ are {\it not} the spherical polar coordinates
because ${\cal F}$ depends on $\vartheta$. The dipole equator ${\cal F},_r = 0$
is at $\cot (\vartheta_{\rm eq}/2) = S/S_F$ (so $\vartheta_{\rm eq} = \pi/2$ in
the Friedmann region including the QSS boundary, see further in this section).
At $r = r_F$ we have ${\cal F} = 1$ and $(\vartheta, \varphi)$ become the
spherical coordinates with the origin at $r = 0$.

Along a geodesic we denote
\begin{equation}\label{3.6}
\left(k^t, k^r, k^{\vartheta}, k^{\varphi}\right) \df \dr {(t, r, \vartheta,
\varphi)} {\lambda},
\end{equation}
where $\lambda$ is an affine parameter. In the $(\vartheta, \varphi)$
coordinates, the geodesic equations for (\ref{3.3}) -- (\ref{3.4}) are
\begin{equation}
\dr {k^t} {\lambda} + \frac {{\cal N} {\cal N},_t} {1 + 2E} \left(k^r \right)^2
+ \frac{\Phi {\Phi,_t}}{{\cal F}^2} \left[\left(k^{\vartheta}\right)^2 + \sin^2
\vartheta \left(k^{\varphi}\right)^2\right] = 0, \label{3.7}
\end{equation}
\begin{eqnarray}
\dr {k^r} {\lambda} &+& 2 \frac {{\cal N},_t} {\cal N} k^t k^r \nonumber \\
&+& \left(\frac {{\cal N},_r} {\cal N} - \frac {E,_r} {1 + 2E}\right)
\left(k^r\right)^2 + 2 \frac {S,_r \sin \vartheta \Phi} {S {\cal F}^2 {\cal N}}\
k^r k^{\vartheta} \nonumber \\
&-& \frac {\Phi (1 + 2E)} {{\cal F}^2 {\cal N}}
\left[\left(k^{\vartheta}\right)^2 + \sin^2 \vartheta
\left(k^{\varphi}\right)^2\right] = 0, \label{3.8} \\
\dr {k^{\vartheta}} {\lambda} &+& 2 \frac {\Phi,_t} {\Phi} k^t k^{\vartheta} -
\frac {S,_r \sin \vartheta {\cal N}} {S \Phi (1 + 2E)}\ \left(k^r\right)^2 + 2
\frac {\cal N} {\Phi} k^r k^{\vartheta} \nonumber \\
&+& \frac {{\cal F},_{\vartheta}} {\cal F}\ \left[- \left(k^{\vartheta}\right)^2
+ \sin^2 \vartheta \left(k^{\varphi}\right)^2\right] \nonumber \\
&-& \cos \vartheta \sin \vartheta \left(k^{\varphi}\right)^2 = 0, \label{3.9} \\
\dr {k^{\varphi}} {\lambda} &+& 2 \frac {\Phi,_t} {\Phi} k^t k^{\varphi} + 2
\frac {\cal N} {\Phi} k^r k^{\varphi} \nonumber\\
&+& 2 \left[\frac {\cos \vartheta} {\sin \vartheta} - \frac {{\cal
F},_{\vartheta}} {\cal F}\right] k^{\vartheta} k^{\varphi} = 0. \label{3.10}
\end{eqnarray}
The geodesics determined by (\ref{3.7}) -- (\ref{3.10}) are null when
\begin{equation}\label{3.11}
\left(k^t\right)^2 - \frac {{\cal N}^2 \left(k^r\right)^2} {1 + 2E(r)} -
\left(\frac {\Phi} {\cal F}\right)^2 \left[\left(k^{\vartheta}\right)^2 + \sin^2
\vartheta \left(k^{\varphi}\right)^2\right] = 0.
\end{equation}

On past-directed rays $k^t < 0$, and $\lambda$ along each of them can be chosen
such that at the observation point
\begin{equation}\label{3.12}
k^t_o = -1.
\end{equation}
(On future-directed rays $k^t > 0$ and a convenient choice of $\lambda$ is
$k^t_e = +1$.)

For correspondence with Ref. \cite{Kras2016a}, in the Friedmann region we choose
the coordinates so that
\begin{equation}\label{3.13}
S = \sqrt{a^2 + {r_F}^2} = S_F.
\end{equation}
Then, throughout the Friedmann region, ${\cal F} = 1$ and $(\vartheta, \varphi)$
are the spherical coordinates. They coincide with the coordinates of the QSS
region at $r = r_F$.

To calculate $k^r$ on nonradial rays, (\ref{3.11}) will be used, which is
insensitive to the sign of $k^r$. This sign will be changed at each point where
$k^r$ reaches zero by the numerical program integrating \{(\ref{3.7}),
(\ref{3.9}) -- (\ref{3.11})\}.

Note that $\vartheta \equiv 0$ and $\vartheta \equiv \pi$ are solutions of
(\ref{3.9}). These rays intersect every space of constant $t$ on the symmetry
axis; they are called axial rays. As follows from (\ref{3.9}), there exist no
null geodesics on which $k^{\varphi} \equiv 0$ and $\vartheta$ has any constant
value other than 0 or $\pi$ (because with $0 \neq \vartheta \neq \pi$,
$k^{\varphi} \equiv 0$ and $k^{\vartheta} = 0$ at a point, Eq. (\ref{3.9})
implies $\dril {k^{\vartheta}} {\lambda} \neq 0$). Consequently, in the axially
symmetric case the only analogues of radial directions are $\vartheta = 0$ and
$\vartheta = \pi$; along these geodesics $\varphi$ is undetermined.

Along a ray emitted at $P_e$ and observed at $P_o$, with $k^{\alpha}$ being
affinely parametrised, we have
\begin{equation}\label{3.14}
1 + z = \frac {\left(u_{\alpha} k^{\alpha}\right)_e} {\left(u_{\alpha}
k^{\alpha}\right)_o},
\end{equation}
where $u_{\alpha}$ are four-velocities of the emitter and of the observer
\cite{Elli1971}. In our case, both the emitter and the observer comove with the
cosmic matter, so $u_{\alpha} = {\delta^0}_{\alpha}$, and the affine parameter
is chosen so that (\ref{3.12}) holds; then
\begin{equation}\label{3.15}
1 + z = - {k_e}^t.
\end{equation}
Equation (\ref{3.10}) has the first integral:
\begin{equation}\label{3.16}
k^{\varphi} \sin^2 \vartheta \Phi^2 / {\cal F}^2 = J_0,
\end{equation}
where $J_0$ is constant along each geodesic. When (\ref{3.16}) is substituted in
(\ref{3.11}), the following results:
\begin{equation}\label{3.17}
(k^t)^2 = \frac {{\cal N}^2 \left(k^r\right)^2} {1 + 2E} + \left(\frac {\Phi}
{\cal F}\right)^2 \left(k^{\vartheta}\right)^2 + \left(\frac {J_0 {\cal F}}
{\sin \vartheta \Phi}\right)^2.
\end{equation}
At the observation/emission point, (\ref{3.12})/(\ref{3.15}), respectively,
apply. Equations (\ref{3.17}) and (\ref{3.15}) show that for rays emitted at the
BB, where $\Phi = 0$, the observed redshift is infinite when $J_0 \neq 0$. A
necessary condition for infinite blueshift ($1 + z_o = 0$) is thus $J_0 = 0$, so

(a) either $k^{\varphi} = 0$,

(b) or $\vartheta = 0, \pi$ along the ray (note that (\ref{3.16}) implies
$J_0/\sin \vartheta \to 0$ when $\vartheta \to 0, \pi$).

\noindent Condition (b) appears to be also sufficient, but so far this has been
demonstrated only numerically in concrete examples of QSS models
(\cite{Kras2016b,Kras2017}).

Condition (a) is {\em not} sufficient, and Ref. \cite{Kras2016b} contains
numerical counterexamples: there exist rays that proceed in a surface of
constant $\varphi$, but approach the BB with $z \to \infty$; the value of
$\vartheta$ along them changes and is different from $0, \pi$. For those rays,
(\ref{3.17}) with the last term being zero implies one more thing
\begin{eqnarray} \label{3.18}
&& {\rm If\ } \lim_{t \to t_B} z = \infty\ {\rm and}\ \lim_{t \to t_B}
\left|k^r\right| < \infty \nonumber \\
&& {\rm then}\ \lim_{t \to t_B} k^{\vartheta} = \pm \infty,
\end{eqnarray}
i.e., such rays approach the BB tangentially to the surfaces of constant $r$.
Examples will appear in Sec. \ref{drift}.

Consider a ray proceeding from event $P_1$ to $P_2$ and then from $P_2$ to
$P_3$. Denote the redshifts acquired in the intervals $[P_1, P_2]$, $[P_2, P_3]$
and $[P_1, P_3] = [P_1, P_2] \cup [P_2, P_3]$ by $z_{12}$, $z_{23}$ and
$z_{13}$, respectively. Then, from (\ref{3.14}),
\begin{equation} \label{3.19}
1 + z_{13} = \left(1 + z_{12}\right) \left(1 + z_{23}\right).
\end{equation}
Thus, for a ray proceeding to the past from $P_1$ to $P_2$, and then back to the
future from $P_2$ to $P_1$:
\begin{equation}\label{3.20}
1 + z_{12} = \frac 1 {1 + z_{21}}.
\end{equation}

\section{The Extremum Redshift Surface}\label{ERS}

\setcounter{equation}{0}

Let a null geodesic stay in the surface $\{\vartheta, \varphi\} = \{\pi, {\rm
constant}\}$. Then $k^r \neq 0$ at all its points, see (\ref{3.11}). Assume it
is past-directed and has its initial point at $r = 0$. Thus, $r$ has to increase
along it and can be used as a parameter. Using (\ref{3.15}), we then obtain from
(\ref{3.7})
\begin{equation}\label{4.1}
\dr z r = \frac {{\cal N} {\cal N},_t} {1 + 2E}\ k^r.
\end{equation}
Since ${\cal N} \neq 0$ from no-shell-crossing conditions \cite{HeKr2002} and
$k^r > 0$, the extrema of $z$ on such a geodesic occur where
\begin{equation}\label{4.2}
{\cal N},_t \equiv \Phi,_{tr} - \Phi,_t {\cal F},_r/{\cal F} = 0.
\end{equation}
Since $\vartheta = \pi$ was assumed, the set defined by (\ref{4.2}) is
2-dimensional; it is the Extremum Redshift Surface (ERS) \cite{Kras2016b}.
Equation (\ref{4.2}) is equivalent to \cite{Kras2016b}
\begin{equation}\label{4.3}
\chi^4 + \chi^3 = - k^3 \left[\frac {r t_{B,r}} {4 M_0 \left(1 - r S,_r /
S\right)}\right]^2,
\end{equation}
where
\begin{equation}\label{4.4}
\chi \df \sinh^2 (\eta/2)
\end{equation}
With $k < 0$, (\ref{4.3}) is solvable for $\chi$ at any $r$, since its left-hand
side is independent of $r$, monotonic in $\chi$ and varies from 0 to $+\infty$,
while the right-hand side is non-negative. The right half of the ERS profile
with the parameters of (\ref{2.20}) is shown in Fig. \ref{realhump}.

In the limit $S,_r \to 0$ (which occurs at $a \to \infty$), (\ref{4.3})
reproduces the equation of the Extremum Redshift {\em Hyper}surface (ERH) of
Ref. \cite{Kras2016a}.

Equation (\ref{4.3}) was derived for null geodesics proceeding along $\vartheta
= \pi$, where the mass dipole is maximum, ${\cal F},_r/{\cal F} = S,_r/S > 0$.
With $S$ given by (\ref{2.21}) we have
\begin{equation}\label{4.5}
F_1 \df 1/\left(1 - rS,_r/S\right) = (r/a)^2 + 1 > 1,
\end{equation}
so, at a given $r$, the ERS has a greater $\eta$ (greater $t - t_B$) than the
ERH of the corresponding L--T model. Moreover, the $\chi$ of (\ref{4.3}) is
greater when $a$ is smaller.

Conversely, for a ray proceeding along the dipole minimum (where $\vartheta =
0$), the factor $F_1$ is replaced by
\begin{equation}\label{4.6}
F_2 \df 1/\left(1 + rS,_r/S\right) = \frac {a^2 + r^2} {a^2 + 2r^2} < 1,
\end{equation}
and so the ERS has a {\em smaller} $t - t_B$ than the ERH in L--T. Also here, a
smaller $a$ has a more pronounced effect.

Extrema of redshift exist along other directions than $\vartheta = 0$ and
$\vartheta = \pi$ (see examples in Sec. \ref{durGRB}), but a general equation
defining their loci remains to be derived.

\section{The time-dependent deflection of light rays}\label{deflect}

\setcounter{equation}{0}

From here on we will consider only the rays proceeding within the surface of
constant $\varphi$.

\begin{figure}[h]
\hspace{-1cm} \includegraphics{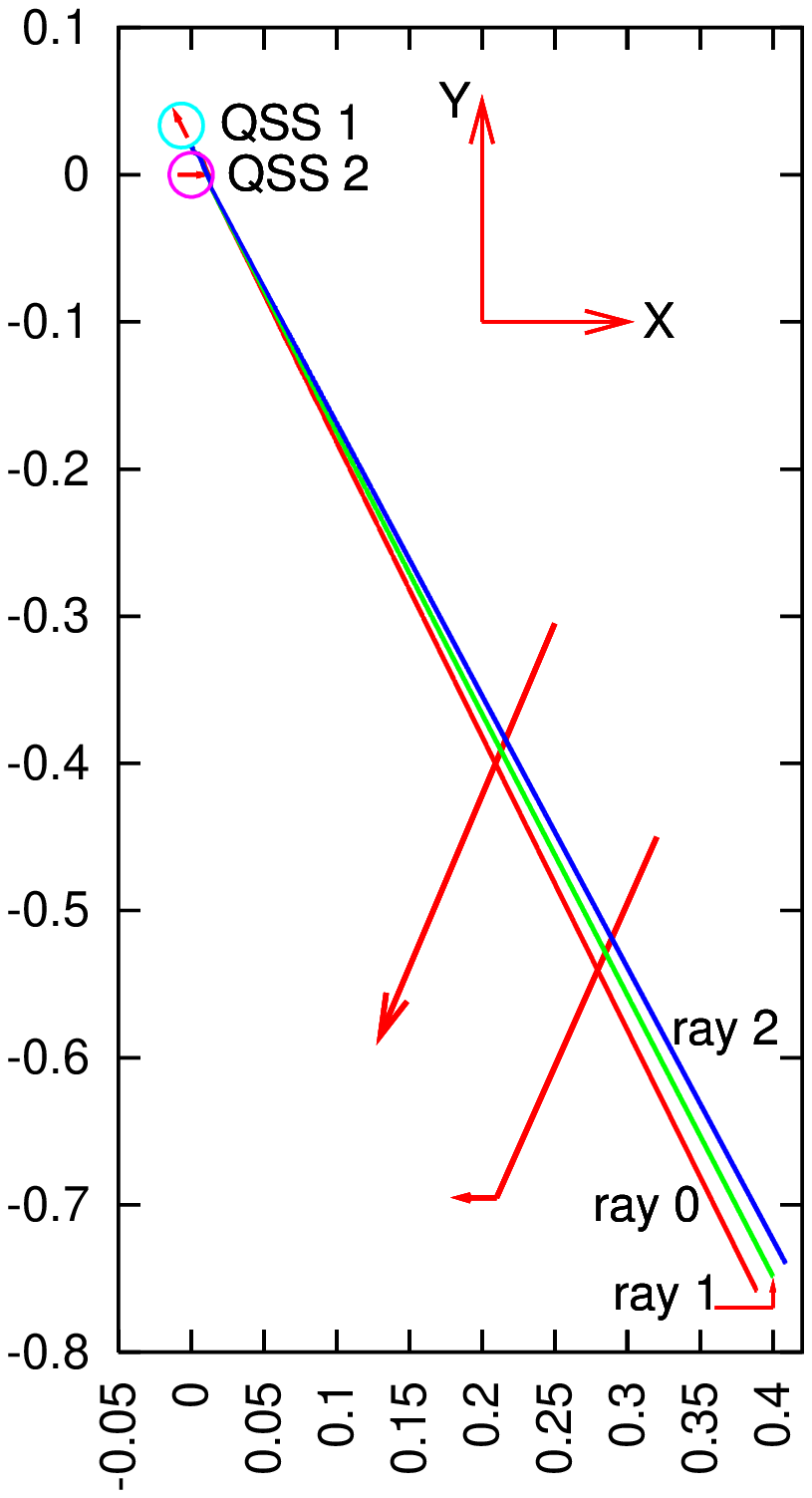}
 ${  }$ \\ [-7.7cm]
\hspace{-2.4cm} \includegraphics[scale=0.42]{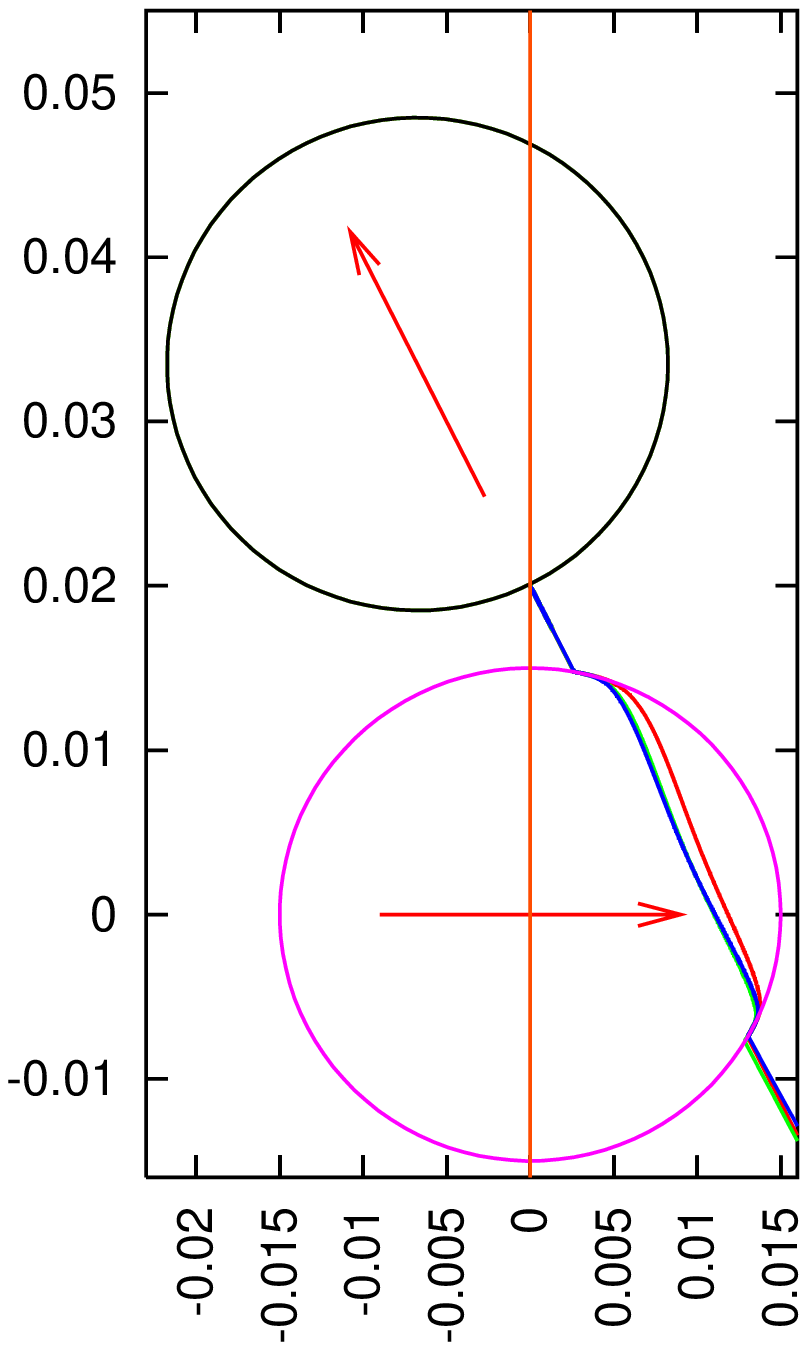}
 ${  }$ \\ [-8.975cm]
\hspace{3cm} \includegraphics[scale=0.3]{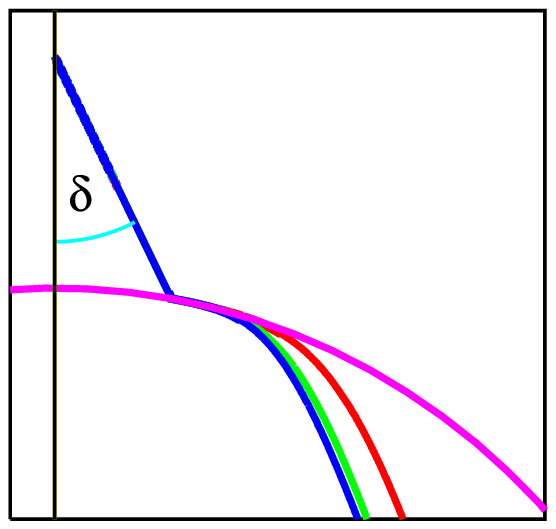}
 ${  }$ \\ [-1.5mm]
\hspace{5.5cm} \includegraphics[scale=0.3]{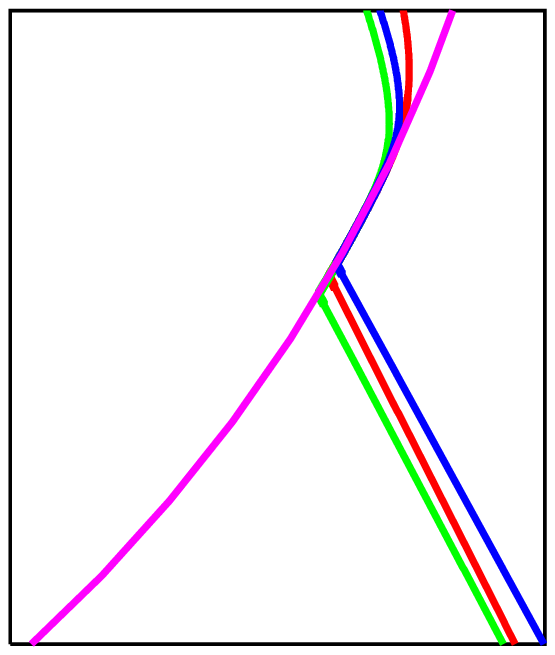}
 ${  }$ \\ [6.5cm]
\caption{Projection of Rays 0, 1 and 2 on a surface of constant $t$ and
$\varphi$. The coordinates are related to those of (\ref{3.3}) -- (\ref{3.4}) by
$X = - r \cos \vartheta, Y = r \sin \vartheta$ ($\vartheta$ increases
clockwise). QSS1 is the Szekeres region that generates the blueshift, QSS2
deflects the rays. The arrows inside circles mark the directions of mass dipole
maxima. The rays end at the present time, $t = 0$. The insets show the
neighbourhoods of the entry and exit points of the rays to the QSS2 region. More
explanation in the text. }
 \label{uprec}
\end{figure}

For the source of the gamma rays we take the BB hump of Fig. \ref{uprec}. We
assume that the maximally blueshifted axial ray emitted there (at the big dot in
Fig. \ref{realhump} and near to the tip of the arrow in the QSS1 circle in Fig.
\ref{uprec}) passes, on its way to the observer, through another axially
symmetric QSS region (QSS2 in Fig. \ref{uprec}), where it is deflected. The
parameters of QSS2 are assumed the same as in QSS1. In general, the angle of
deflection changes with time in consequence of the cosmic drift
\cite{KrBo2011,KoKo2017}. Consequently, if the maximally blueshifted ray hits a
given observer at a given time, an axial ray emitted a while later will miss the
observer, apart from a few exceptional cases (see below). After a short while,
the rays seen by the observer in the same direction will be coming from other
regions of the same source and will be less blueshifted, i.e., they will form
the afterglow. See Sec. \ref{drift} for a more detailed description.

A ray that hits QSS2 along the symmetry axis (from either the dipole maximum or
minimum side) is not deflected \cite{KrBo2011}. But the directions in the dipole
equator plane, $\vartheta = \pi/2$, are also special. If a ray hits QSS2 there
having proceeded along $\vartheta = \pi/2$ in the Friedmann region, then a
deflection will occur, but there will be no drift.\footnote{This effect was
verified numerically; its reason is not evident in the geodesic equations.} For
rays that hit QSS2 from other directions, the time-dependent deflection is to be
expected.

In the following, we will use two coordinate charts of the class (\ref{3.3}) --
(\ref{3.5}), one centered at the origin of QSS1, and the other centered at the
origin of QSS2. Their $t$ will be the same, but the $r$-coordinates will differ.
Whenever confusion may arise, we will denote the $r$ of QSS1 by $r_{\rm S1}$,
and the other one by $r_{\rm S2}$.

We consider a ray that was emitted from the LSH at the symmetry axis of QSS1 on
the dipole maximum side, and proceeds along this axis. The initial point is the
one that gives the maximally strong blueshift. (It is ``ray A'' in Sec. 9 of
Ref. \cite{Kras2017} and ``the ray'' shown in Fig. \ref{realhump}). The ray then
passes all through QSS1 and emerges into the Friedmann region (after passing the
origin of QSS1 it becomes ``ray B'' of Ref. \cite{Kras2017}). This part of the
ray's path is not shown in Fig. \ref{uprec}; in the following it will be called
the Upray. On it, we choose the point just behind the QSS1 region, with
\begin{eqnarray}
&& (r_{\rm S1}, t) = (r, t)_1 = \label{5.1} \\
&& (0.0150907023847052114, -0.13926900571845113), \nonumber \\
&& z_1 = -0.39590497158301252 \label{5.2}
\end{eqnarray}
(these numbers are taken from the numerical tables for ray B \cite{Kras2017}).
We take $(t_1, z_1)$ as the initial data for Ray 0. (Note: $z_1$ is the {\em
upward} redshift between $r_{\rm S1} = 0$ and $r_{\rm S1} = r_1$. The true
(downward) redshift between these points is $1 + z_{1\ {\rm true}} = 1 / (1 +
z_1)$.)

For the initial $r_{\rm S2}$ on Ray 0 we take
\begin{equation}\label{5.3}
r_{\rm S2} = r_i = 0.02.
\end{equation}
We assume that the dipole maximum of QSS2 lies on $\vartheta = \pi$ in Fig.
\ref{uprec}, while the angle between the direction of the ray at $(t_1, r_i)$
and the dipole equator of QSS2 is
\begin{equation}\label{5.4}
\delta = \arctan(1/2);
\end{equation}
see the uppermost inset in Fig. \ref{uprec}. The values of $r_i$ and $\delta$
were chosen such that the drift is clearly visible in the illustrations, while
the two QSS regions are not too close to each other. The angle of deflection
becomes larger when $r_i$ is smaller and vice versa. The angle $\delta$ is
related to the initial $k^{\vartheta}$ by
\begin{equation}\label{5.5}
\sin \delta = \frac {k^{\vartheta}_i \Phi(t_i, r_i)} {1 + z_1}
\end{equation}
(see derivation in Appendix \ref{rayangle}), with $\Phi(t_i, r_i)$ calculated
from (\ref{2.5}) using (\ref{2.12}) and (\ref{2.13}), and (\ref{2.17}) for
$t_B$.

The upward redshift to $t = 0$ on Ray 0 is\footnote{For reproducibility of the
results, the numbers are quoted up to 17 decimal digits. Such precision was
needed to capture time differences of $\approx 10$ min at the observer -- see
Secs. \ref{drift} -- \ref{durGRB}.}
\begin{equation}\label{5.6}
(1 + z)_{\rm 3\ up} = 7.19256480334251602 \times 10^{-3}.
\end{equation}
The $z_1$ in (\ref{5.2}) is the upward $z$ between $r_{\rm S1} = 0$ and $r_{\rm
S1} = r_1$. Consequently, (\ref{5.6}) is the upward $z$ between $r_{\rm S1} = 0$
and the present observer. The proper $z$ between the LSH and $r_{\rm S1} = 0$
was calculated in Ref. \cite{Kras2017}:
\begin{equation}\label{5.7}
1 + z_{\rm ols3} = 1.11939135405414447 \times 10^{-7}.
\end{equation}
So, the proper $1 + z$ between the last scattering in QSS1 and the present time
is
\begin{equation}\label{5.8}
1 + z_3 = \frac {1 + z_{\rm ols3}} {1 + z_{\rm 3\ up}} = 1.55631737031266357
\times 10^{-5}.
\end{equation}
This is near to $1.553 \times 10^{-5}$ of Ref. \cite{Kras2017}; the difference
arose because in Ref. \cite{Kras2017} the intervening QSS2 region was absent.
The $1 + z_3$ of (\ref{5.8}) is within the range $1 + z < 1.689 \times 10^{-5}$
needed to blueshift the emission frequencies of hydrogen into the gamma sector
\cite{Kras2016a}.

Ray 0 overshot the present time $t = 0$ in consequence of numerical errors. The
calculation stopped at
\begin{equation}\label{5.9}
t_{\rm fin} = 1.29738987343870121 \times 10^{-10}\ {\rm NTU},
\end{equation}
and the other coordinates of the endpoint were
\begin{eqnarray}\label{5.10}
&& (r_{\rm S2}, \vartheta) = (r, \vartheta)_{\rm fin} = \\
&& (0.85217686701400219,\ 4.2386753628314251) \nonumber
\end{eqnarray}
(recall: $\vartheta$ is counted from the $X < 0$ direction clockwise).

Rays 1 and 2 have their initial points at the same $r_{\rm S2} = r_i = 0.02$ as
Ray 0, and the initial direction still determined by $\delta$ of (\ref{5.4}),
but the initial $t$ later by 0.00002 NTU (Ray 1) and 0.00004 NTU (Ray 2). The
angle $\delta$ being the same means that the initial directions are
parallel-transported along the world line of the emitter; see Appendix
\ref{ParTrans} for more on this.

As seen in Fig. \ref{uprec}, each ray is deflected by a different angle when
passing through QSS2. This is the cosmic drift (non-repeatability of light
paths) of Refs. \cite{KrBo2011,QABC2012,KoKo2017}. On comparing the two upper
insets in Fig. \ref{uprec} one sees that the projections of Rays 0 and 2
intersect at the edge of QSS2: Ray 0 goes further than Ray 2 around the BB hump.
Similarly, the projections of Ray 0 and Ray 1 intersect at point P of
coordinates
$(X, Y) \approx (0.019987, -0.02143)$.

\begin{figure}[h]
\hspace{-7mm} \includegraphics[scale=0.65]{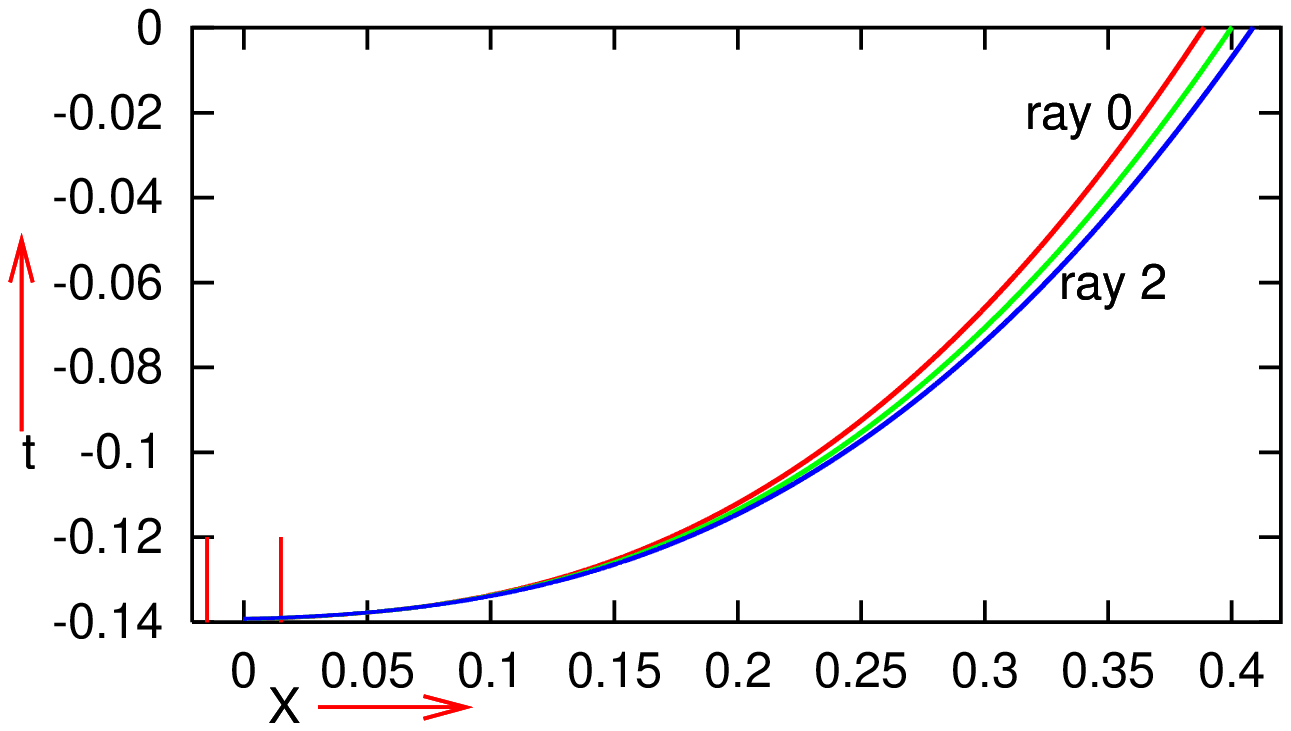}
 ${  }$ \\ [-4.8cm]
\hspace{-2.3cm} \includegraphics[scale=0.3]{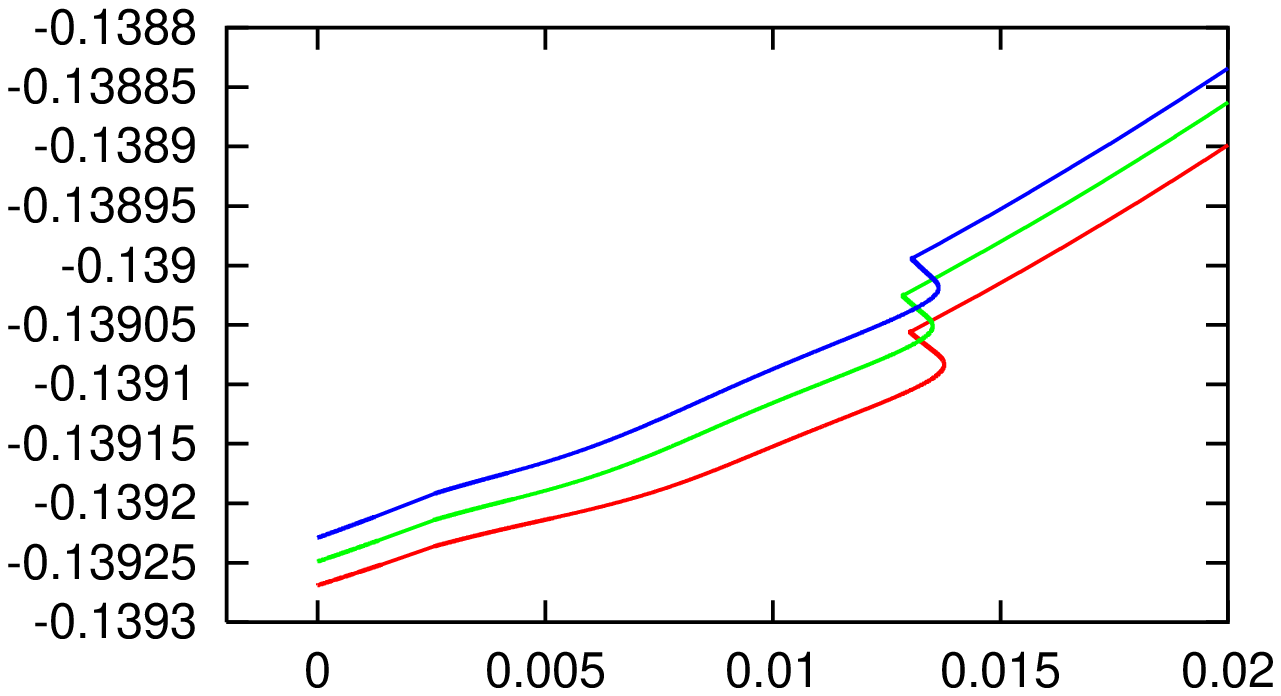}
 ${  }$ \\ [2cm]
\caption{Projection of rays 0 -- 2 on the $Y = 0$ coordinate plane. The vertical
strokes show the borders of QSS2. The inset shows the segments of the rays
inside and near QSS2.}
 \label{uptx}
\end{figure}

\begin{figure}[h]
\hspace{-4mm} \includegraphics[scale=0.5]{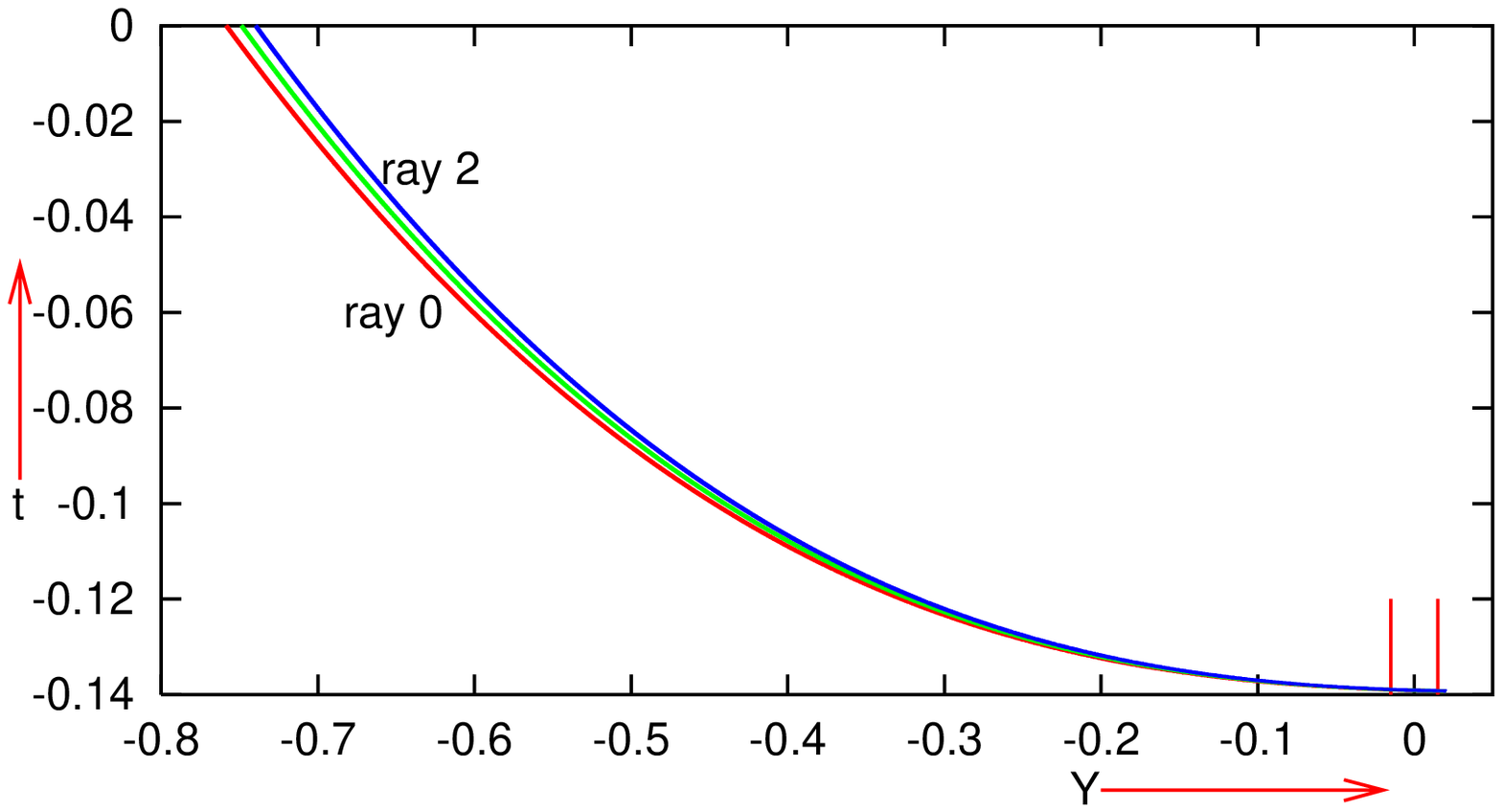}
 ${  }$ \\ [-4.2cm]
\hspace{3cm} \includegraphics[scale=0.3]{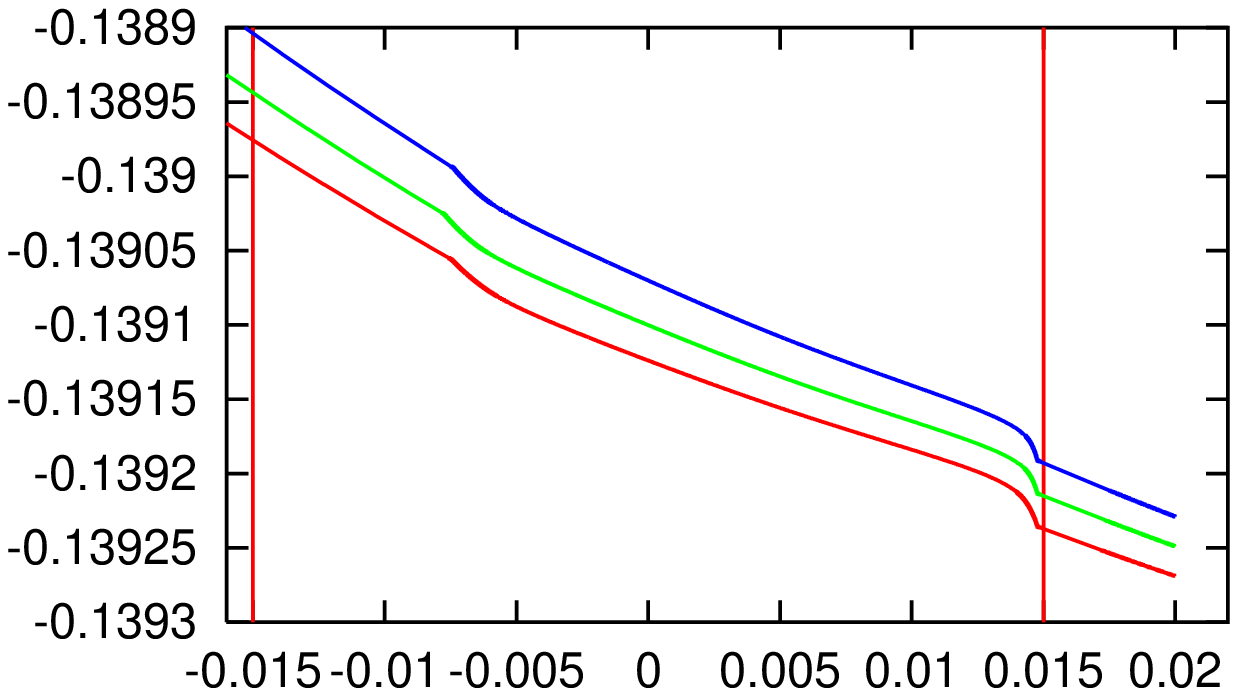}
 ${  }$ \\ [2cm]
\caption{Projection of rays 0 -- 2 on the $X = 0$ coordinate plane. The vertical
strokes show the borders of QSS2. The inset shows the segments of the rays
inside and near QSS2.}
 \label{upty}
\end{figure}

Although the projections of Rays 0 and 1 on the $(X, Y)$-surface intersect at P,
the rays do not intersect in spacetime, as is seen from their projections on the
$(X, t)$ and $(Y, t)$ coordinate surfaces in Figs. \ref{uptx} and \ref{upty},
respectively. The intersecting $(X, Y)$-projections imply that an observer at P
would register both Ray 0 and (suitably later, from a different direction) Ray
1.

\begin{figure}[h]
\hspace{-3mm} \includegraphics[scale=0.7]{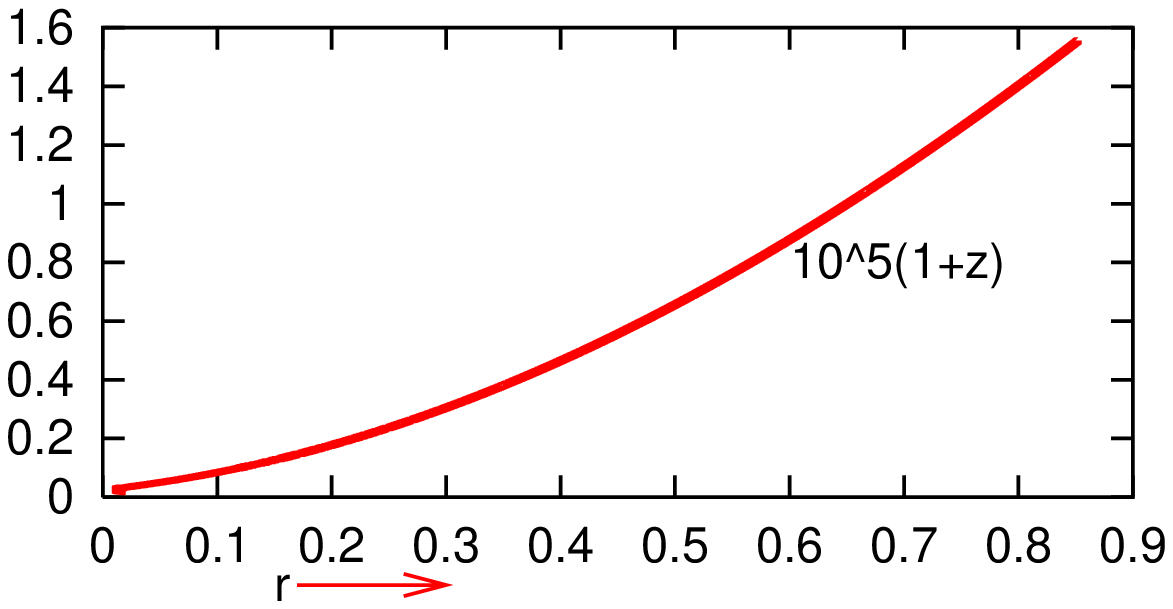}
 ${  }$ \\ [-4cm]
\hspace{-3.2cm} \includegraphics[scale=0.3]{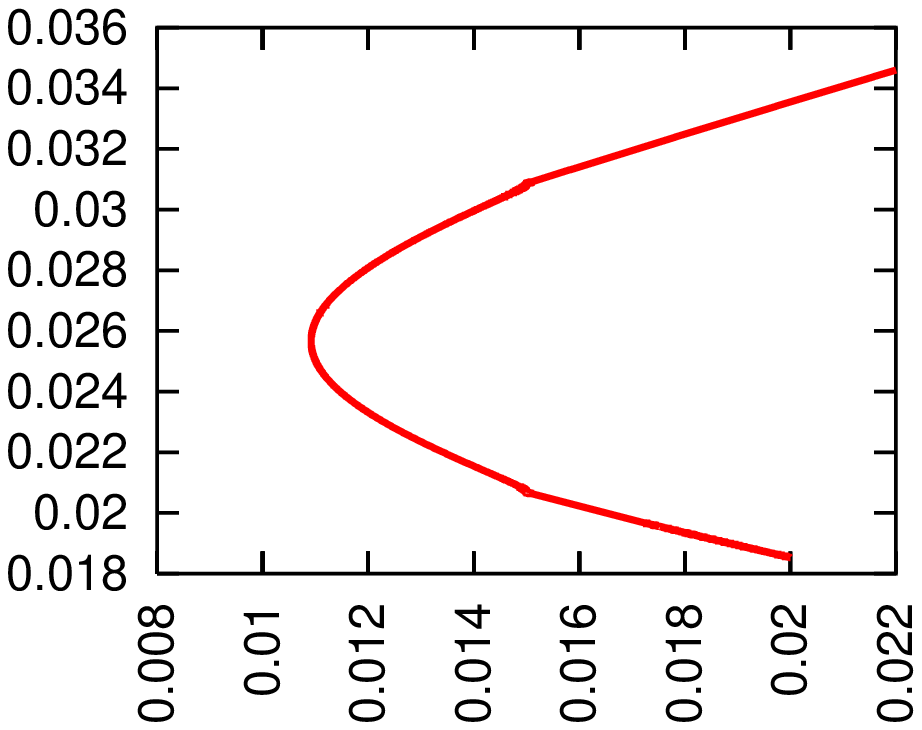}
 ${  }$ \\ [2cm]
 \caption{{\bf Main panel:} The graph of $[1 + z(r)] \times 10^5$ along Ray 0.
{\bf Inset:} The segment of the graph where the local extrema of $z(r)$ are
present; they are displayed in Fig. \ref{upzrlupa}.}
 \label{upzr}
\end{figure}

\begin{figure}[h]
\includegraphics[scale=0.4]{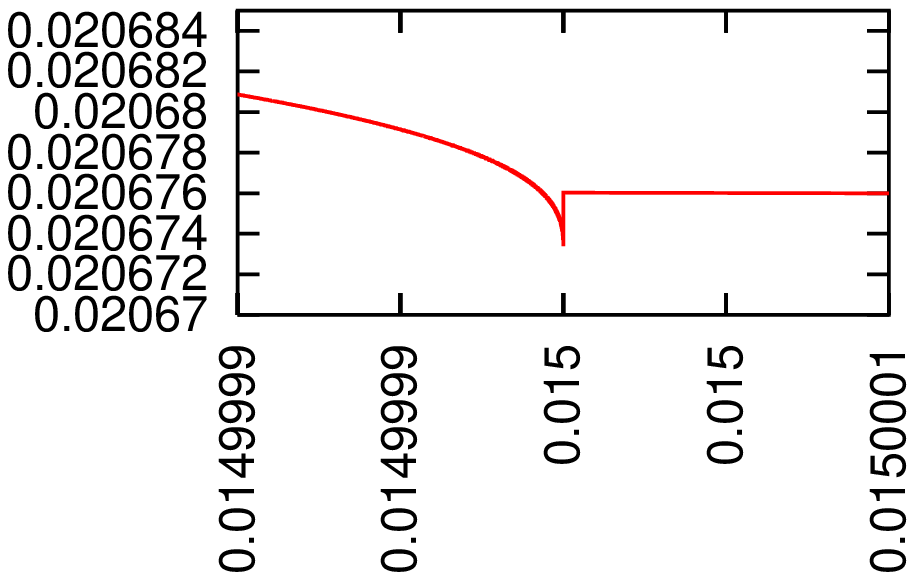}
\includegraphics[scale=0.4]{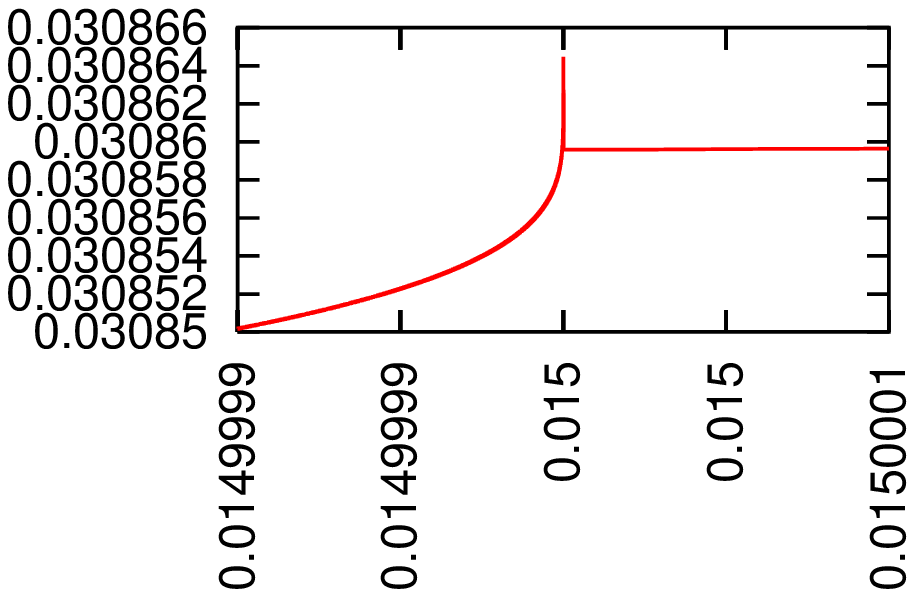}
\caption{Closeup views on the curve from Fig. \ref{upzr} around the points where
it crosses the edges of QSS2.}
 \label{upzrlupa}
\end{figure}

Figures \ref{upzr} and \ref{upzrlupa} show $[1 + z(r)] \times 10^5$ along Ray 0.
In short segments near the borders of QSS2, $1 + z$ decreases when followed
along the ray to the future. However, the times of flight of the rays through
QSS2 are made longer by the deflections. The net result is that the ray spends
more time in the redshift-generating region, and the final $1 + z$ is larger
than it would be in the absence of QSS2 -- see (\ref{5.8}) and the remark under
it.

\section{Rate of angular drift of the rays}\label{drift}

\setcounter{equation}{0}


Now we calculate the average rate of the angular drift of the rays applying a
few approximations. As the first approximation, we take the angles between the
images of the future-end segments of the rays in Fig. \ref{uprec} for the real
angles in spacetime. Since these angles are small, the departure from the exact
result should be tolerable. As shown in Appendix \ref{fininit}, the angle
between the end segments of Rays 2 and 0 is
\begin{equation}\label{6.1}
\Delta \beta = 1.79491638375138503^{\circ}.
\end{equation}
The time difference between rays 0 and 2 at their initial points at $r_{\rm S2}
= r_i$ is, from (\ref{2.14}) and (\ref{2.15}),
\begin{eqnarray}\label{6.2}
&& \Delta t = 0.00004\ {\rm NTU} = 39.2 \times 10^5\ {\rm y} \nonumber \\
&& = 2.06192 \times 10^{12}\ {\rm min}.
\end{eqnarray}
This implies the average rate of angular drift, with respect to the source time
\begin{eqnarray}\label{6.3}
\dr {\beta} t &=& 4.57886832589639038 \times 10^{-7}\ ^{\circ}/{\rm y}
\nonumber \\
&=& 8.70507286292089426 \times 10^{-13}\ ^{\circ}/{\rm min} \nonumber \\
&=& 0.15193218308397861 \times 10^{-13}\ {\rm rad}/{\rm min}.\ \ \ \
\end{eqnarray}
In (\ref{6.3}), the time is counted between the initial points of Rays 0 and 2,
at $(r, \vartheta)_{\rm S2} = (r_i, \pi/2)$, with $r_i$ given by (\ref{5.3}).
The upward redshift on Ray 0 from $r_{\rm S1} = 0$ to the present time is the
$z_{\rm 3\ up}$ given by (\ref{5.6}). Hence, the proper redshift on Ray 0
between $r_{\rm S2} = r_i$ and the present time is
\begin{equation}\label{6.4}
1 + z_{\rm ip} = \frac {1 + z_1} {1 + z_{\rm 3\ up}} = 83.98881969561866230,
\end{equation}
where $z_1$ is given by (\ref{5.2}). So, the time difference $\Delta t$ at
$(r_i, t_1)$ results in the time difference $(1 + z_{\rm ip}) \times \Delta t$
at the observer. The rate of angular drift at the observer is thus
\begin{eqnarray}\label{6.5}
&& \left.{\dr {\alpha} t}\right|_{\rm observer} = \left.\dr {\beta} t\right/
\left(1 + z_{\rm ip}\right) \nonumber \\
&& = 1.80895723543432844 \times 10^{-16}\ {\rm rad/min}.
\end{eqnarray}
As a curiosity, in Ref. \cite{KrBo2011} (an off-center observer in a spherical
void, observing galaxies around her) the cosmic drift was $\leq 10^{-6}$
arcsec/y $\approx 9.22 \times 10^{-18}$ rad/min, i.e. $\approx$ 19.6 times
smaller than (\ref{6.5}).

The observed GRBs typically last from less than a second to a few minutes
\cite{Perlwww}. Let 10 minutes be the reference time. Imagine a ray in our model
reaching the observer 10 minutes later than Ray 0 from the same direction. Will
it still be a gamma ray?

A direct answer would require emitting a ray from the observer position back in
time later by
\begin{equation}\label{6.6}
\tau = 10\ {\rm min} = 1.93993947388841468 \times 10^{-16}\ {\rm NTU}
\end{equation}
(from (\ref{2.14}) -- (\ref{2.15})) than the $t_{\rm fin}$ of (\ref{5.9}), in
the direction opposite to Ray 0. But a ray having its initial point so close to
the endpoint of Ray 0 could be numerically distinguishable from Ray 0 only at
very high precision that is inaccessible to this author.\footnote{Such a ray
would travel through nearly the whole lifetime of the Universe. To capture time
differences counted in seconds, the numerical time-step would have to be of the
order of 1 s. Using (\ref{2.14}) and taking $13.819 \times 10^9$ y for the age
of the Universe \cite{Plan2014}, the calculation would require $\approx 43
\times 10^{16}$ steps.} Therefore, we shall again resort to an approximate
estimate.

Ray 2 was sent from its initial point later by 0.00004 NTU than Ray 0. This
translates, by (\ref{6.4}), to
\begin{equation}\label{6.7}
\Delta t_2 \approx 0.00335955278782475\ {\rm NTU}
\end{equation}
at the present time, $t = 0$. So, we send a ray (call it Backray 0) from the
observer position given by (\ref{5.9}) -- (\ref{5.10}) backward in time, with
the initial $t = t_{\rm fin} + \Delta t_2$, and with the initial direction
opposite to the final direction of Ray 0. There were difficulties with assigning
the correct number to the initial direction, see Appendix \ref{misalign}. The
number finally chosen was
\begin{equation}\label{6.8}
\mu \df k^{\vartheta}/k^r = 0.0107412585641537794
\end{equation}
With this value, at the first contact with the edge of QSS2 Backray 0 and Ray 0
coincide to better than $10^{-8}$ in both $X$ and $Y$; see Appendix
\ref{misalign} again (but the coincidence is not as good all the way).

\begin{figure}[h]
\includegraphics[scale=0.8]{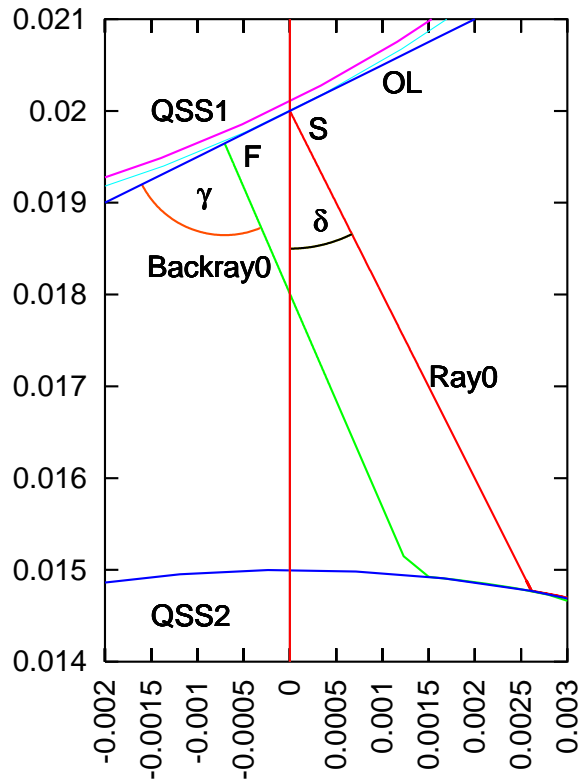}
\caption{Projections of Ray 0 and Backray 0 on a surface of constant $t$ in a
vicinity of QSS1. The line OL is orthogonal to the projection of Ray 0 at its
initial point S; Backray 0 was stopped where its projection crossed OL, at point
F in spacetime. The thin arc tangent to OL at S is concentric with QSS1 and has
the radius $r_1$ given by (\ref{5.1}). The other arcs are borders of the QSS
regions.}
 \label{backraylupa}
\end{figure}

The integration of Backray 0 was stopped at the 2-surface $S_b$ in spacetime
given by the equations
\begin{eqnarray}\label{6.9}
r &=& \frac {r_i} {\sin \vartheta + \tan \delta \cos \vartheta} \equiv \frac
{r_i \cos \delta} {\sin (\delta + \vartheta)}, \nonumber \\
\varphi &=& {\rm constant}.
\end{eqnarray}
This surface is orthogonal to Ray 0 at its initial point S. It intersects the
plane of Fig. \ref{backraylupa} along the straight line OL, which has the
equation
\begin{equation}\label{6.10}
Y = r_i + X \tan \delta,
\end{equation}
where $r_i$ and $\delta$ are given by (\ref{5.3}) -- (\ref{5.4}) and $(X, Y)$
are the coordinates in Fig. \ref{backraylupa}. The point where Backray 0 hit
$S_b$ will be denoted F. Stopping Backray 0 there was convenient for prolonging
it into the QSS1 region.\footnote{The discontinuity in the slope of Backray 0 at
the edge of QSS2 in Fig. \ref{backraylupa} resulted from interpolation -- only 1
in 600 calculated points is used by the drawing program. \label{brline}}

The surface $S_b$ and the line OL are at the same time tangent at S to the
circle of radius $r_{\rm S1} = r_1$, where Ray 0 took over from Upray; see Fig.
\ref{backraylupa}.

The following parameters of Backray 0 at F will be needed for further
calculations:
\begin{eqnarray}
r_b &=& 0.0196597503597857326, \label{6.11} \\
t_b &=& -0.13922666531999014, \label{6.12} \\
\vartheta_b &=& 1.5348829837350006, \label{6.13} \\
1 + z_b &=& 74.080313337692317, \label{6.14} \\
{\cal R}(t_b) &=& 0.00617900818769279159. \label{6.15}
\end{eqnarray}

Imagine a family of past-directed rays, all emitted from $(r, \vartheta) = (r,
\vartheta)_{\rm fin}$ of (\ref{5.10}), in the direction opposite to the final
direction of Ray 0, with their emission instants continuously increased from $t
= t_{\rm fin}$ of (\ref{5.9}) to $t_{\rm fin} + \Delta t_2$ of (\ref{6.7}). Call
this family ${\cal G}$ (for ``geodesics''). The earliest of these rays is the
time-reverse of Ray 0, the last coincides with Backray 0. Their intersections
with the surface $S_b$ form an arc connecting point S to point F. The arc SF
represents the spatial drift of the endpoint of Backray 0 with respect to the
initial point of Ray 0.

The angle $\gamma$ between the tangent vector to the arc SF at F and the
direction vector of Backray 0 at F is
\begin{eqnarray}\label{6.16}
\gamma &=& 1.5137231069990575\ {\rm rad} \nonumber \\
&=& 86.729945382475933^{\circ} \quad \Longrightarrow
\nonumber \\
\widetilde{\gamma} &\df& \pi/2 - \gamma = 0.0570732197958390142\ {\rm rad}
\nonumber \\
&=& 3.2700546175240777^{\circ}.
\end{eqnarray}
see Appendix \ref{angamma}. The $(X, Y)$ coordinates of F are
\begin{eqnarray}\label{6.17}
X_b &=& - r_b \cos \vartheta_b = -7.05895596023793498 \times 10^{-4}, \nonumber \\
Y_b &=& r \sin \vartheta_b = 0.0196470734618825471,
\end{eqnarray}
where $r_b$ and $\vartheta_b$ are given by (\ref{6.11}) and (\ref{6.13}). The
time-coordinates of points S and F in spacetime differ by
\begin{equation}\label{6.18}
\Delta t_1 = t_b - t_1 = 4.234039846099 \times 10^{-5}\ {\rm NTU}.
\end{equation}

Now take a ray of family ${\cal G}$ (call it Backray N) that was emitted back in
time from $(r, \vartheta)_{\rm fin}$ by $U \Delta t_2$ later than the $t_{\rm
fin}$ of (\ref{5.9}), where $U$ is small and $\Delta t_2$ is given by
(\ref{6.7}). The point of intersection of Backray N with the surface $S_b$,
denoted $P_N$, lies on the arc SF near to the point S. Quantities referrring to
$P_N$ will carry the subscript N. We make two more approximations:

1. The arc $SP_N$ coincides with the straight line segment $S\overline{P_N}$,
tangent to $SP_N$ at S, where $\overline{P_N}$ is the projection of $P_N$ on the
plane of Fig. \ref{backraylupa}.

2. The $X_N$ coordinate of $P_N$, the angle $\widetilde{\gamma}_N$ between
Backray N and the normal to the arc $SP_N$ at $P_N$, the $\Delta t_N \df t_N -
t_1$ and the redshift difference $z_{\rm ip} - z_N$ are related to $X_b$,
$\widetilde{\gamma}$, $\Delta t_1$ and $z_{\rm ip} - z_b$ all by the factor $U$.

It is not obvious how precise these approximations are because, as was seen in
Fig. \ref{uprec}, the directions of the rays do not necessarily change
uniformly: the projections of some rays intersect. We leave the problem of
verifying the correctness of these assumptions (and possible improvements upon
them) for the future. The approximation should be better, the shorter
$S\overline{P_N}$ is.

We wish to calculate by how much the past endpoint $P_N$ and the direction of
Backray N at $P_N$ will differ from those of Ray 0 at S when $U = 10\ {\rm
minutes} / \Delta t_2$, i.e.,
\begin{equation}\label{6.19}
U =  5.77439795236701905 \times 10^{-14}
\end{equation}
from (\ref{6.6}) and (\ref{6.7}). Then we obtain at $P_N$
\begin{eqnarray}
X_N &\df& -U r_b \cos \vartheta_b \nonumber \\
&=& -4.07612208426468965 \times 10^{-17}, \label{6.20} \\
Y_N &\df& r_i + X_N / \tan \delta = r_i + 2 X_N \nonumber \\
&=& 1.99999999999999185 \times 10^{-2}, \label{6.21} \\
\widetilde{\gamma}_N &=& U \widetilde{\gamma} \nonumber \\
&=& 3.29563483524085620 \times 10^{-15}\ {\rm rad},\ \ \ \ \label{6.22} \\
\Delta t_N &=& U \Delta t_1  \label{6.23} \\
&=& 2.44490310175544341 \times 10^{-18}\ {\rm NTU}, \nonumber \\
1 + z_N &=& 1 + z_{\rm ip} - U \left(z_{\rm ip} - z_b\right) \nonumber \\
&=& 83.988819695618090. \label{6.24}
\end{eqnarray}
{}From (\ref{6.20}) -- (\ref{6.21}), the $(r, \vartheta)_{\rm S2}$ coordinates
of $P_N$ are
\begin{eqnarray}
\vartheta_N &=& \arctan \left[Y_N/\left(- X_N\right)\right] \nonumber \\
&=& 1.5707963267948946\ {\rm rad}, \label{6.25} \\
r_N &=& Y_N / \sin \vartheta_N \nonumber \\
&=& 1.99999999999999185 \times 10^{-2}. \label{6.26}
\end{eqnarray}
The program that calculated this does not see the difference between $Y_N$ and
$r_N$. This is because $\pi/2 - \vartheta_N = 2.03806104213235315 \times
10^{-15}$, and the Fortran program finds $\sin \vartheta_N = 1$ up to 16th
decimal place. (The difference $\pi/2 - \vartheta_N$ is not to be confused with
$\widetilde{\gamma}_N$. The $\vartheta_N$ is the value of $\vartheta_{\rm S2}$
at the endpoint of Backray N, while $\widetilde{\gamma}_N$ is the angle between
the direction of Backray N at the endpoint and the normal to ${\overline {\rm
SF}}$ there.)

The point $\overline{P_N}$, lying on OL very near to the point S, lies very
nearly on the circle of radius $r_{\rm S1} = r_1$ surrounding the QSS1 region.
So, we make one more approximation and assume that $\overline{P_N}$ lies on that
circle. The error in $r$ committed thereby is smaller than $10^{-17}$.

\section{Accounting for the short-livedness of the gamma-ray flash}\label{durGRB}

\setcounter{equation}{0}

We now send a new ray (call it Downray 1), backward in time, in prolongation of
Backray N. What is the blueshift or redshift on this ray when it crosses the
LSH?

\begin{figure}[h]
 ${  }$ \\ [3mm]
 \hspace{-1cm}
 \includegraphics[scale=0.65]{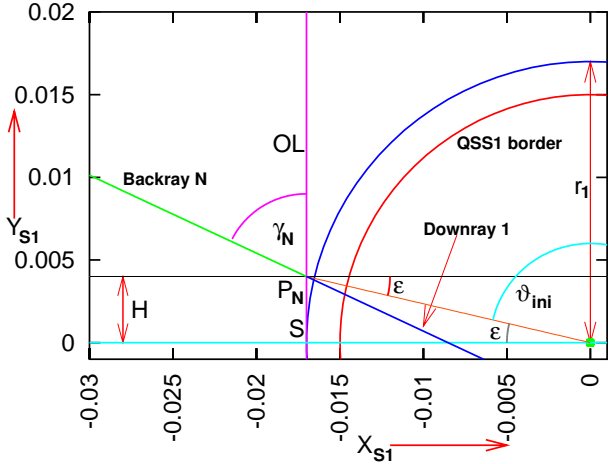}
\caption{Sketch of Backray N near its past endpoint and of Downray 1 that
continues in the same direction into the QSS1 region. The values of $H$, $r_1$,
$\varepsilon$ and $\pi/2 - \gamma_N = \widetilde{\gamma}_N$ are much
exaggerated. The origin of the $(X, Y)$ coordinates in this graph coincides with
the origin of QSS1. }
 \label{sketch}
\end{figure}

The numbers in (\ref{6.20}) and (\ref{6.22}) allow us to calculate the initial
direction for Downray 1 in terms of the $(r, \vartheta)_{\rm S1}$ coordinates;
see the sketch in Fig. \ref{sketch}. It has its initial point at $r_{\rm S1} =
r_1$ given by (\ref{5.1}), shifted by
\begin{eqnarray}\label{7.1}
H &\df& \left|X_N\right| / \cos \delta = \sqrt{5} \left|X_N\right| / 2 \nonumber
\\
&=& 4.55724303250198597 \times 10^{-17}
\end{eqnarray}
toward $Y_{\rm S1} > 0$ with respect to the symmetry axis. Its initial direction
is inclined at the angle $\gamma_N = \pi / 2 - \widetilde{\gamma}_N$ to the OL
line. The $\vartheta_{\rm S1}$ of its initial point is $\varepsilon$ given by
\begin{equation}\label{7.2}
\tan \varepsilon = H/r_1 = 3.01990120560648060 \times 10^{-15}.
\end{equation}
The Fortran 90 program sees no difference between $\varepsilon$ and $\tan
\varepsilon$ up to the level of $10^{-17}$.

The tangent vector to Downray 1 at the point $(t, r, \vartheta) = (t_1 + \Delta
t_N, r_1, \varepsilon)$ is
\begin{eqnarray}
&& k^r_N = \frac {\sqrt {1 - k {r_1}^2} \left(1 + z_N\right)} {{\cal R}_N {\cal
D}} \nonumber \\
&&\ \ \ \ \ \times \left(- \tan \varepsilon \sin \widetilde{\gamma}_N - \sqrt {1
- k {r_1}^2} \cos \widetilde{\gamma}_N\right),\ \ \ \ \ \ \  \label{7.3} \\
&& k^{\vartheta}_N = \frac {1 + z_N} {r_1 {\cal R}_N {\cal D}} \nonumber \\
&& \ \ \ \ \ \times \left(- \sqrt {1 - k {r_1}^2} \sin \widetilde{\gamma}_N +
\tan \varepsilon \cos \widetilde{\gamma}_N\right),\ \ \ \ \ \ \ \label{7.4} \\
&& {\cal D} \df \sqrt{1 - k{r_1}^2 + \tan^2 \varepsilon}, \quad {\cal R}_N \df
{\cal R}(t_N); \label{7.5}
\end{eqnarray}
see calculation and discussion in Appendix \ref{Downini}. In the numerical
calculations the factor $\left(1 + z_N\right)$ was omitted -- because smaller
values of $k^r$ and $k^{\vartheta}$ result in smaller numerical steps of $r$ and
$\vartheta$ which improves the numerical precision. The actual value of $1 + z$
at the endpoint of Downray 1 was then calculated using (\ref{3.19}).

The initial $t$ on Downray 1 is
\begin{equation}\label{7.6}
t_N = t_1 + \Delta t_N = -0.013926900571845113 \ {\rm NTU}.
\end{equation}
Note that this is the same as $t_1$ of (\ref{5.1}) -- with $\Delta t_N \approx
10^{-18}$, the Fortran program does not see the difference. This confirms the
remark made under (\ref{6.6}) -- if determined by direct numerical calculation,
the path of Backray N would become, at some point along the way,
indistinguishable from that of Ray 0.

Summarizing, the initial point of Downray 1 is $(t, r, \vartheta)_{\rm S1} =
(t_N, r_1, \varepsilon)$, with the numbers given in (\ref{7.6}), (\ref{5.1}) and
(\ref{7.2}), respectively, while its initial direction is determined by
(\ref{7.3}) -- (\ref{7.5}).

\begin{figure}[h]
 \hspace{-1cm} \includegraphics[scale=0.66]{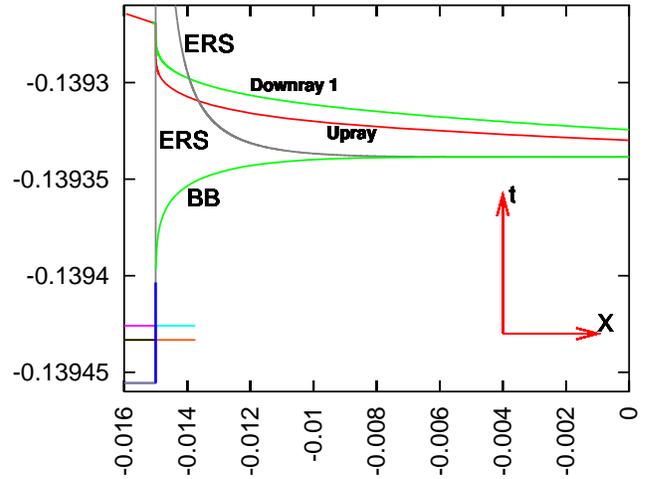}
\caption{Comparison of the $(t, X)$ profiles of Downray 1 and Upray. Explanation
in the text.}
 \label{downuncor}
\end{figure}

\begin{figure}[h]
 \hspace{-6mm} \includegraphics[scale=0.64]{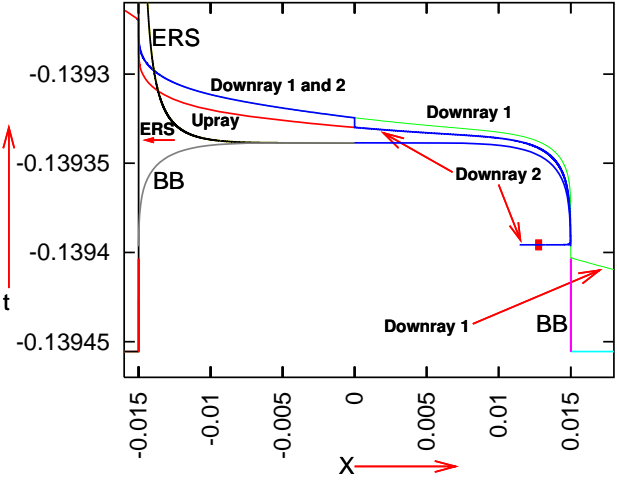}
\caption{The $(t, X)$ profiles of Downray 1, Downray 2 and Upray. The small
square near the lower end of Downray 2 marks its intersection with the last
scattering hypersurface.}
 \label{downcor}
\end{figure}

\begin{figure}[h]
  \includegraphics[scale=0.5]{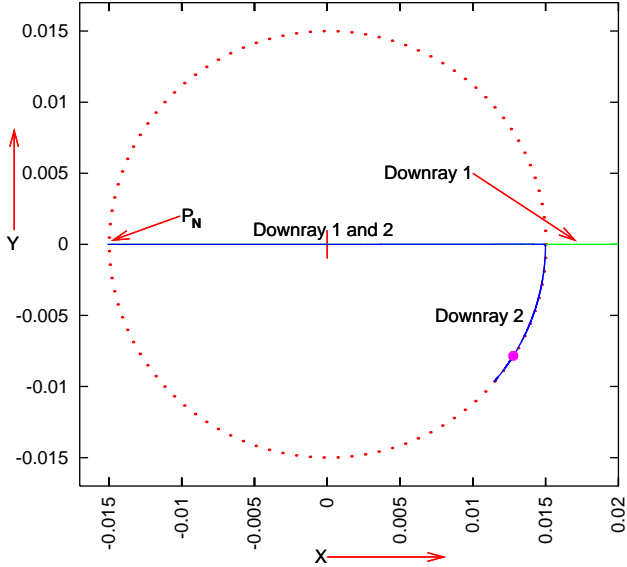}
\caption{Projection of Downrays 1 and 2 on the $(X, Y)$ coordinate plane. The
dot marks the intersection of Downray 2 with the last scattering hypersurface,
the vertical stroke marks the origin at $r_{\rm S1} = 0$. More explanation in
the text.}
 \label{downrays}
\end{figure}

If continued from these initial data without corrections, Downray 1 would have
the $(t, X)$ profile shown in Fig. \ref{downuncor}. The $t$-coordinate along it
disagrees with that of the Upray by much more than should be expected, see
Appendix \ref{inflight} for details. This happens in consequence of a numerical
instability at the outer branch of the ERS, which this author was not able to
overcome. Therefore, somewhere along the way the $t$-coordinate of Downray 1 had
to be hand-corrected. This was done as follows. At the point of closest approach
of Downray 1 to the $r = 0$ line (see Appendix \ref{inflight}), the tangent
vector to the ray was parallel-transported from the actual time calculated along
the ray ($t_o = -0.13932447906821049$ NTU) to the initial time of the Upray,
\begin{equation}\label{7.7}
t'_o = -0.13932991589546649,
\end{equation}
which is by $5.43682725598348959 \times 10^{-6}$ NTU earlier. The parallel
transport was done as in Appendix \ref{ParTrans}: $z$ was thereby unchanged,
while $k^r$ and $k^{\vartheta}$ were multiplied by ${{\cal R} (t_o)}/{{\cal R}
(t'_o)} = 1.3849964189634454$. From there on, the ray was continued as Downray
2, see Figs. \ref{downcor} -- \ref{downrays}. Consequences of other possible
modifications of Downray 1 will be discussed in Sec. \ref{nearby}.

The operation described above was done under two assumptions: (1) that the the
point of closest approach to $r = 0$ and the direction of Downray 1 there were
determined correctly, and (2) that the parallel transport could be done by the
Friedmannian rule of Appendix \ref{ParTrans}.

\begin{figure}[h]
 \hspace{-5mm} \includegraphics[scale=0.5]{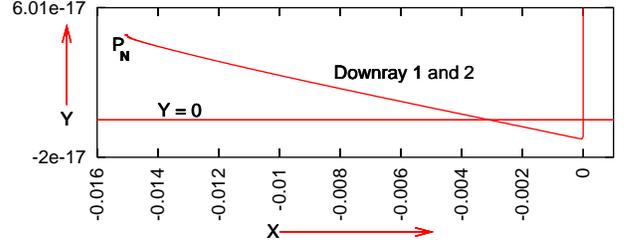}
\caption{Closeup view on Fig. \ref{downrays} in the range $X \in [-0.016,
0.001]$. In this range Downrays 1 and 2 coincide.}
 \label{microray}
\end{figure}

The first assumption is justified by the fact that Downray 1 proceeds close to
the $\vartheta = 0$ axis, with $|Y| < 4.404 \times 10^{-17}$ all the way between
$P_N$ and $X = 0$, see Fig. \ref{microray}. This means that any departure from
this path is smaller than the accuracy of the numerical code.

The second assumption is justified by the fact that at the closest approach to
the origin $r \approx 2.23 \times 10^{-22}$, ${\cal E},_r/{\cal E} \approx 2.23
\times 10^{-19}$ and $\Phi \approx 2.15 \times 10^{-25}$, so the term in
(\ref{2.1}) that is responsible for the difference between the QSS and Friedmann
geometries is $\Phi {\cal E},_r/{\cal E} \approx 4.79 \times 10^{-44}$. The
Friedmann term $\Phi,_r$ is at this location $9.62 \times 10^{-4}$. The
non-Friedmannian correction is thus too small to show up at the level of
accuracy assumed here.

Near the origin some quantities become indeterminate of the type 0/0
\cite{Kras2016b}, so the integration of the ray equations had to include certain
precautions: (1) The numerical step had to be gradually decreased on approaching
the origin. (2) Downray 1 aimed at $r = 0$ with high precision, and at a certain
step its $r$-coordinate shot to $r < 0$. This had to be taken care of. See
Appendix \ref{inflight} for details.

Figure \ref{downcor} shows the $(t, X)$ profiles of Downrays 1 and 2 across the
full diameter of the BB hump. Only the left half of the ERS is shown. Downray 1
at first goes along the steep slope of the BB hump, but then turns away and
crosses the LSH in the Friedmann region at
\begin{equation}\label{7.8}
\left(\begin{array}{l}
t \\
r \\
\vartheta \\
1 + z_P \\
1 + z_o \\
\end{array}\right)_{D1LSH} = \left(\begin{array}{l}
 -0.13945067585653098\ {\rm NTU} \\
 0.0522717572229775690 \\
 3.1415926411632658 \\
 3.9974477620254634 \\
 335.740919327408696 \\
 \end{array}\right)
\end{equation}
($\vartheta_{D1LSH} = 179.999999288^{\circ} \approx \pi$). The $z_P$ is relative
to the point $P_N$, $z_o$ is relative to the present observer calculated from $1
+ z_o = (1 + z_N) (1 + z_P)$ using (\ref{6.24}).

Downray 2 never leaves the QSS1 region and finally hits the BB at the steep
slope -- tangentially to a surface of constant $r$, as predicted in
(\ref{3.18}). It crosses the LSH at the point marked with the small square in
Fig. \ref{downcor} and with the dot in Fig. \ref{downrays}, at
\begin{equation}\label{7.9}
\left(\begin{array}{l}
t \\
r \\
\vartheta \\
1 + z_P \\
1 + z_o \\
\end{array}\right)_{D2LSH} = \left(\begin{array}{l}
 -0.13939573683235620\ {\rm NTU} \\
 0.0149934194406606183 \\
 3.6918768830001061 \\
 0.0268536489146646434 \\
 2.2554062768632145161 \\
 \end{array}\right).
\end{equation}

\begin{figure}[h]
 \hspace{-5mm}  \includegraphics[scale=0.5]{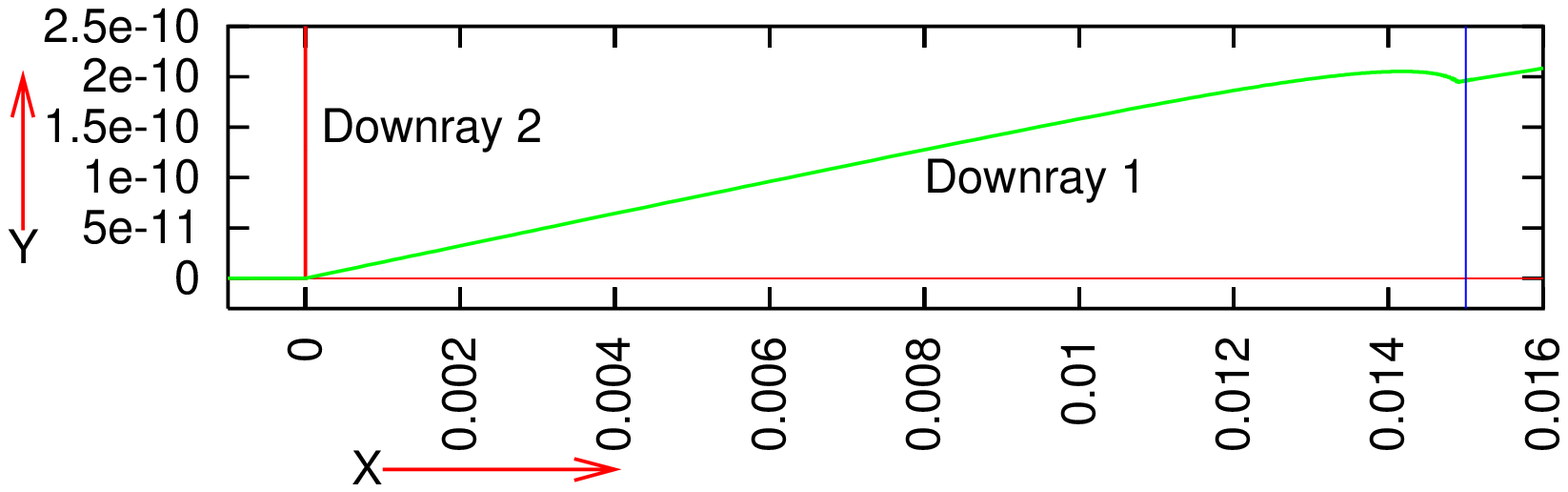}
${ }$ \\ [5mm]
 \hspace{-5mm}  \includegraphics[scale=0.5]{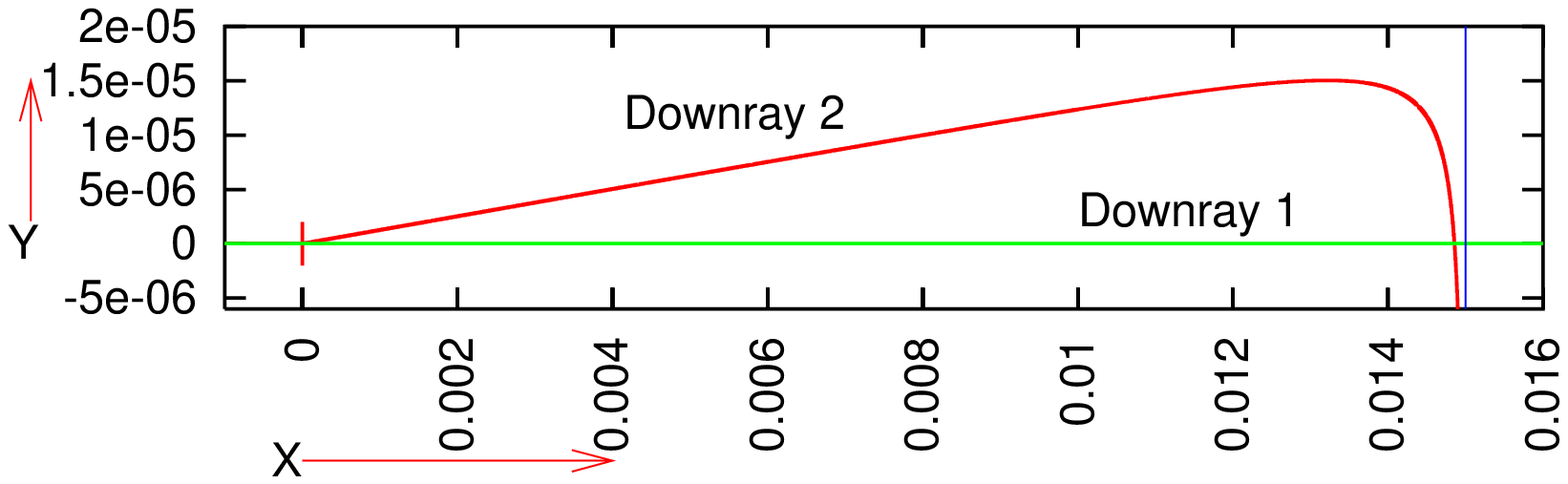}
\caption{Prolongation of Downray 1 and Downray 2 to the range $X \in [-0.001,
0.016]$. Note the different $Y$-scales in the upper and lower panels, and in
Fig. \ref{microray}. The thin vertical line at right is the border of the QSS1
region.}
 \label{microray2}
\end{figure}

Figure \ref{downrays} shows the projection of Downrays 1 and 2 on the $(X, Y)$
coordinate plane. At this scale, both rays seem to proceed across the QSS1
region along the symmetry axis. Figures \ref{microray} and \ref{microray2} show
details of the rays' paths. Downray 1 proceeds along a nearly straight line at
the angle $\varepsilon$ of (\ref{7.2}) to the $Y = 0$ line and intersects it at
$X \approx -0.0030588$. At $(X, Y) \approx (-4.322714809677376 \times 10^{-5}$,
$-1.02236722643458 \times 10^{-17})$ it turns upwards and aims at $r = 0$. After
reaching $(X, Y) \approx (-2.230943471939892 \times 10^{-22},
-1.07257712221974956 \times 10^{-25})$, the Fortran program finds the next $r$
to be negative, see Appendix \ref{inflight} to see how this problem was handled.
On the other side of the origin, the two rays proceed at different angles.
Downray 1 stays within the $0 \leq Y < 2.5 \times 10^{-10}$ strip up to the
boundary of the QSS1 region; see the upper panel of Fig. \ref{microray2}.
Downray 2 is initially inclined to the $Y = 0$ line at an angle that is $\approx
10^{12}$ times greater than in the $X < 0$ sector. On approaching the boundary
of the QSS1 region, it bends around and becomes nearly tangent to a constant-$r$
surface. It continues this way up to the intersection with the BB; see the lower
panel of Fig. \ref{microray2} and Fig. \ref{downrays}.

\begin{figure}[h]
 \hspace{-5mm}  \includegraphics[scale=0.6]{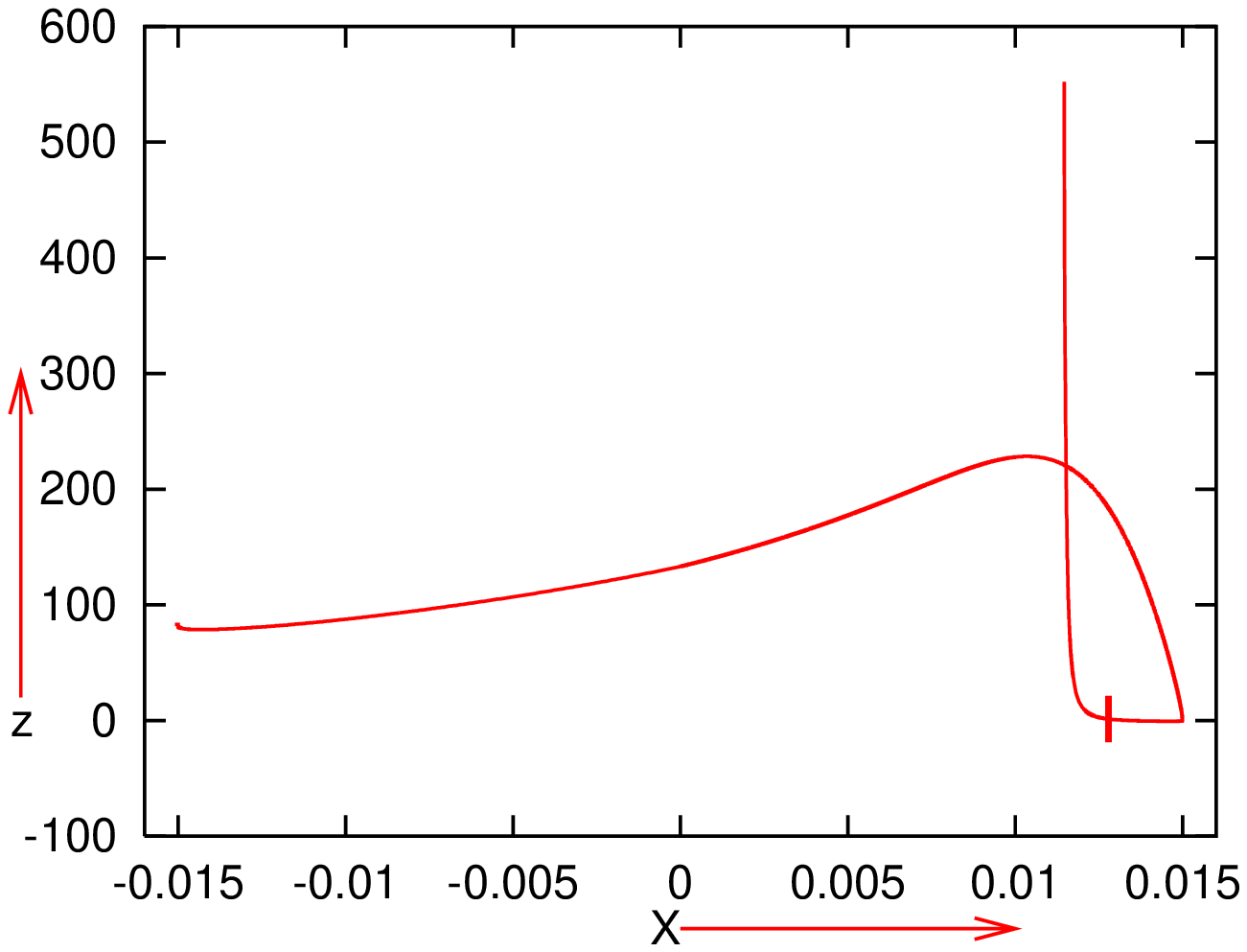}
${ }$ \\ [-5.7cm]
 \hspace{-2.2cm}
 \includegraphics[scale=0.4]{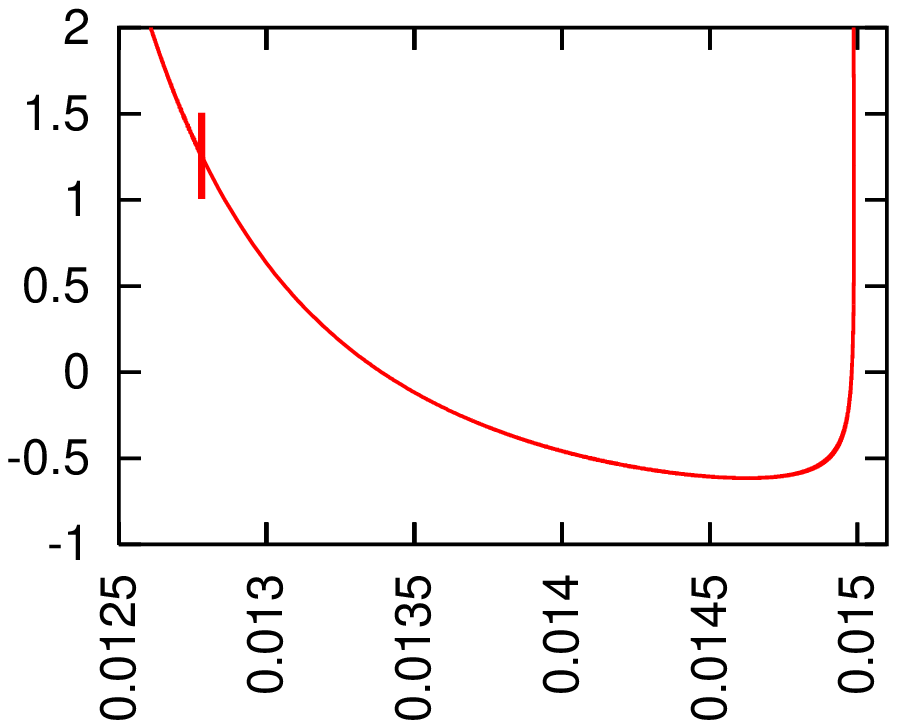}
 ${ }$ \\ [2.7cm]
\caption{The redshift along Downray 2 seen by the present observer. {\bf Inset:}
Closeup view on the neighbourhood of the minimum of $z$. The vertical strokes
mark $z$ at the intersection with the last scattering hypersurface, see the last
line of (\ref{7.9}).}
 \label{realzx}
\end{figure}

Finally, Fig. \ref{realzx} shows the $\{X, z\}$ relation along Downray 2, with
$z$ relative to the present observer. The observer would see blueshift if the
light sources were placed between $X \approx 0.0135$ and $X \approx 0.0149$, but
will not see any blueshift in the light emitted at the LSH. For $z$ at LSH given
by (\ref{7.9}), and with the parameters of the spatially flat $\Lambda$CDM model
($H_0 = 69.6$ km/(s $\times$ Mpc), $\Omega_m = 0.286$, $\Omega_{\Lambda} =
0.714$ \cite{calculator}) the Cosmology Calculator of Refs.
\cite{wripa,calculator} says the source of Downray 2 would be 8.726 Gyr to the
past of the present observer. In the model used here, it is $\approx 0.139$ NTU
= 13.622 Gyr to the past. With $1 + z_o$ given by (\ref{7.9}), the whole range
of visible light is moved into near ultraviolet.

Assuming that the approximations were all valid, Eqs. (\ref{7.9}) and
(\ref{5.8}), and Fig. \ref{realzx} prove that 10 minutes after seeing a
gamma-ray flash the present observer will see radiation coming from the same
direction with frequency $(1 + z_o)_{D2LSH} / (1 + z_3) \approx 1.449 \times
10^5$ times smaller. This will be in the ultraviolet range.

\section{Perturbations of Downray 2}\label{nearby}

\setcounter{equation}{0}

Since the numerical calculations involved approximations, and they proceeded at
the borderline of numerical accuracy, it is useful to investigate the redshift
along rays propagating in proximity of Downray 2. For each ray described below,
the numerical step on approaching the minimum $r$ and the jump through $r = 0$
were handled in the same way as described in Appendix \ref{inflight}.

Four sets of rays were considered:

1. Downrays 3 and 4 had the same initial $(t, r, \vartheta = \varepsilon)$ and
the same initial direction as Downray 2, but the corrected $t$ at closest
approach to $r = 0$ larger (ray 3) and smaller (ray 4) than the $t'_o$ of
(\ref{7.7}).

In the next three sets, the corrected $t$ at closest approach to $r = 0$ was
(\ref{7.7}).

2. Downrays 5 -- 7 had the same initial $(t, r, \vartheta = \varepsilon)$ as
Downray 2, but $\widetilde{\gamma}_N$ zero, minus that of Downray 2 and twice
that of Downray 2, respectively ($\widetilde{\gamma_N} = 0$ means that the
initial direction is parallel to the $Y = 0$ line).

3. Downrays 8 -- 10 had the same initial $(t, r)$ as Downray 2, but
$\varepsilon$ twice as large, and the same values of $\widetilde{\gamma}_N$ as
Downrays 5, 6 and 2, respectively.

4. Downrays 11 -- 13 had the same initial $(t, r)$ as Downray 2, but half the
value of $\varepsilon$, and the same set of values of $\widetilde{\gamma}_N$ as
in point 3.

For Downray 3 trial time-intervals $\tau_n = 10^{-n}$ NTU were added to the
$t'_o$ of (\ref{7.7}) to verify the effect. With $n = 18$ the effect was
invisible, as expected. With $n = 17, 16, \dots, 8$ the changes in $(t, r,
\vartheta, 1 + z_P)$ at the LSH compared to those of Downray 2 were small and
qualitatively insignificant. Only at $n = 7$ a substantial change occurred. The
ray hit the BB at the boundary of the QSS1 region at earlier $t$ than Downray 2,
but the program could not detect its intersection with the LSH. The ray with
$\tau_{7.2} = 2 \times 10^{-7}$ NTU (Downray 3 in the figures) flew over the BB
hump and intersected the LSH in the Friedmann region, with $(r, 1 + z_o) \approx
(0.03724, 8.0629)$.

For Downray 4, the trial intervals added to $t'_o$ were $\widetilde{\tau}_n = -
10^{-n}$ NTU. Similarly to Downray 3, a noticeable difference occurred only at
$n = 6$: the ray hit the LSH well before reaching the edge of the QSS1 region,
with $(r, 1 + z_P) \approx (0.0145, 182.256)$.

The conclusion is that the value of $1 + z$ at the LSH is not sensitive to the
values of $\tau_n$ and $\widetilde{\tau}_n$ as long as $\tau \leq 10^{-8}$ NTU =
980 years, i.e., it is not sensitive to numerical errors smaller than $10^{-8}$
NTU.

Details of the other results are described in Appendix \ref{apnearby}. Here is
just a short summary:

On rays 5 to 9 and 12, $1 + z_o$ was sufficient to move the whole visible light
range into the ultraviolet; see Table \ref{zondownrays} in Appendix
\ref{apnearby} and Ref. \cite{emspec} for the numbers. On Downray 10, the
visible range was moved into the near ultraviolet just adjacent to the visible
range. On rays 11 and 13, the visible range was slightly shifted toward violet
(Downray 11) or toward red (Downray 13), but most of it remained visible.

The final conclusion is that the result calculated in Sec. \ref{durGRB} remains
valid for past-directed rays that reach the neighbourhood of the point $P_N$
with directions close to that of Downray N: on all those rays, the present
observer would see redshift or only weak blueshift ($1 + z \approx 0.91$ on
Downray 11). A strong blueshift $1 + z \approx 10^{-5}$ would only be seen on
Ray 0 that goes along the symmetry axis of the QSS1 region.

In fact, this result was to be expected on the basis of the comments to eq.
(\ref{3.17}), but it is instructive to have it confirmed by explicit
computations.

\section{Conclusions and prospects}\label{sumup}

\setcounter{equation}{0}

The present paper is a continuation and development of Ref. \cite{Kras2017}. In
that paper, it was shown that if a light ray is emitted at the last-scattering
hypersurface in an axially symmetric quasispherical Szekeres region (denoted
QSS1) along the symmetry axis, with the frequency in the visible range, then it
will follow this axis all through QSS1 and might be blueshifted by so much that
the present observer will see it as a gamma ray. The axial direction is
unstable, and rays proceeding in a close neighbourhood of the symmetry axis will
be now observed in the ultraviolet range.

In the present paper, it was investigated what happens to the maximally
blueshifted ray (MBR) if, on its way between the emission point and the present
observer, it passes through another QSS region (denoted QSS2). Both QSS regions
are matched into a Friedmann background of negative spatial curvature, and the
observer is in this background.

The QSS metrics and some of their properties were introduced in Secs. \ref{QSSS}
-- \ref{ERS}. In Sec. \ref{deflect}, it was shown that if the MBR passes through
QSS2 along a generic path,\footnote{Non-generic paths are along the symmetry
axis and in the plane of the Szekeres dipole equator.} then is deflected by a
continuously changing angle via the cosmic drift effect
\cite{KrBo2011,KoKo2017}. Therefore, any observer can find herself in this ray's
path only at a single instant of time. At all other times, the rays coming from
the same direction will have lower frequencies (i.e. will be redshifted or only
slightly blueshifted), while the MBR will miss the observer. The rate of angular
drift of the MBR was calculated in Sec. \ref{drift}. Using this result, in Sec.
\ref{durGRB} it was demonstrated by a numerical calculation that an MBR that
reached a present observer at some instant will no longer be visible to her
after 10 minutes. Instead, the ray coming from the same direction will be in the
ultraviolet range. Finally, in Sec. \ref{nearby} (with numerical details in
Appendix \ref{apnearby}) it was shown that the result of Sec. \ref{durGRB} is
not sensitive to perturbations smaller than $10^{-8}$ NTU = 980 years of the
instant when the rays cross the origin of the QSS1 region, and smaller than
$\approx 3.3 \times 10^{-15}$ rad $= 1.89 \times 10^{-13}\ ^{\circ}$ of their
direction on exit from the QSS1 region.

In the present model the gamma-ray flash is instantaneous. As with any model,
this is an idealisation. In reality, mechanisms should exist that will stretch
the instant of observation of the gamma ray to a short but finite interval.
Modelling them may require using more general metrics and other BB profiles than
those used here. Refs. \cite{Kras2016a,Kras2016b,Kras2017} and the present paper
provided a proof of existence of a mechanism that can shift the range of visible
light frequencies to the gamma range, and that will ensure short-livedness of
the high-frequency flash.

The two QSS regions were assumed axially symmetric for simplicity. Then the
geodesic equations show that the MBR must follow the symmetry axis all through
the QSS1 region, so in the numerical calculations it can be kept exactly on this
path, even though it is unstable \cite{Kras2016b,Kras2017}. But strong
blueshifts are generated also by fully nonsymmetric QSS regions
\cite{Kras2016b}. In the nonsymmetric case, the MBR will be changing its
direction with time in consequence of the cosmic drift within the source
\cite{KrBo2011,KoKo2017}, even in the absence of the deflecting region. This
process is a more difficult numerical challenge: the maximally blueshifted ray
must be identified by numerics alone \cite{Kras2016b}.

The model presented here does not solve the problem of too-long-lasting
afterglows. This will be the subject of a separate paper.

\appendix

\section{Calculating $\delta$ in (\ref{5.5})}\label{rayangle}

\setcounter{equation}{0}

The unit vector tangent to the line $(t, \vartheta, \varphi) = (t_1, \pi/2, {\rm
constant})$ in the Friedmann region is, using (\ref{3.3}),
\begin{equation}\label{a.1}
n^{\alpha}_2 = \left(0, \frac {\sqrt{1 + 2E}} {\cal N}, 0, 0\right).
\end{equation}
The unit direction vector of a null vector $k^{\alpha}$ is \cite{Elli1971}
\begin{equation}\label{a.2}
n^{\alpha}_1 = u^{\alpha} - \frac {k^{\alpha}} {k^{\rho} u_{\rho}}.
\end{equation}
In our case $u^{\alpha} = \delta^{\alpha}_0$. Consequently, at the initial point
of the ray, using (\ref{3.15}), eq. (\ref{a.2}) implies
\begin{equation}\label{a.3}
n^{\alpha}_1 = \left(0, \frac {k^r_i} {1 + z_1},  \frac {k^{\vartheta}_i} {1 +
z_1}, 0\right),
\end{equation}
where $z_1$ is given by (\ref{5.2}) (account was taken of $k^{\varphi} = 0$).
{}From (\ref{3.17}) with $J_0 = 0$ we have, again using (\ref{3.15}),
\begin{equation}\label{a.4}
k^r_i = \pm \frac {\sqrt{1 + 2E_i}} {{\cal N}_i} \sqrt{\left(1 + z_1\right)^2 -
\left(\frac {\Phi_i k_i^{\vartheta}} {{\cal F}_i}\right)^2}.
\end{equation}
So, the angle $\delta$ between $n^{\alpha}_1$ and $n^{\alpha}_2$ is given by
\begin{equation}\label{a.5}
\cos \delta = - g_{\alpha \beta} n^{\alpha}_1 n^{\beta}_2 = \sqrt{1 -
\left[\frac {k^{\vartheta}_i \Phi_i} {{\cal F}_i \left(1 +
z_1\right)}\right]^2}.
\end{equation}
Since the initial point of Ray 0 is in the Friedmann region, where ${\cal F}
\equiv 1$, the above is equivalent to (\ref{5.5}).

\section{Parallel transport along $u^{\alpha}$ in a Friedmann
region}\label{ParTrans}

\setcounter{equation}{0}

In a Friedmann region a vector $k^{\mu}$ parallel-transported along $u^{\alpha}
= {\delta_0}^{\alpha}$ obeys $k^t =$ constant, $k^i {\mathcal R} =$ constant, $i
= 1, 2, 3$, i.e., the ratio $k^{\vartheta}/k^r$ stays constant. Then
\begin{eqnarray}
k^t_{\ell t} &=& k^t_{et}, \label{b.1} \\
(k^r, k^{\vartheta})_{e t} &=& \frac {{\cal R} (t_{\ell t})} {{\cal R} (t_{e
t})}\ (k^r, k^{\vartheta})_{\ell t}, \label{b.2}
\end{eqnarray}
where $\ell t$ and $et$ denote ``at a later time'' and ``at an earlier time'',
respectively. If $k^{\mu}$ is a null vector and we wish to reset $z$ from
$z_{\ell t}$ to $z_{e t}$, then (\ref{b.2}) is modified as follows:
\begin{equation}\label{b.3}
(k^r, k^{\vartheta})_{e t} = \frac {{\cal R} (t_{\ell t})} {{\cal R} (t_{e t})}\
(k^r, k^{\vartheta})_{\ell t} \times \frac {1 + z_{et}} {1 + z_{\ell t}}.
\end{equation}

\section{Calculating the angles for (\ref{6.1}) -- (\ref{6.5})}\label{fininit}

\setcounter{equation}{0}

The direction of Ray 0 in the $(X, Y)$ surface at the future endpoint can be
read off from the table of values of $X$ and $Y$ found by the same program that
calculated the path of the ray. The last two pairs in this table are
\begin{eqnarray}\label{c.1}
&& (X, Y)_1 \nonumber \\
&& = (0.38875803859076119, -0.75833536622992936), \nonumber \\
&& (X, Y)_2 \\
&& = (0.38875804873013375, -0.75833538643402199). \nonumber
\end{eqnarray}

Hence the final direction of Ray 0 in the $(X, Y)$ surface is inclined to the $Y
= 0$ line at the angle $\beta_0$, where
\begin{eqnarray}\label{c.2}
\left|\tan \beta_0\right| &\df& \left|\dr X Y\right|_{r_{\rm fin}} = \frac {X_2
- X_1} {\left|Y_2 - Y_1\right|} \nonumber \\
&=& 0.50184745960551459, \nonumber \\
&\Longrightarrow& \beta_0 = 26.64966989421893587^{\circ}.
\end{eqnarray}
In the same way we find for Ray 2:
\begin{equation}\label{c.3}
\beta_2 = 28.44458627797032091
\end{equation}
So, $\Delta \beta = \beta_2 - \beta_0$ is as given by (\ref{6.1}).

\section{Correcting the direction of Backray 0 in Sec.
\ref{durGRB}}\label{misalign}

\setcounter{equation}{0}

\begin{figure}[h]
 \hspace{-4cm} \includegraphics[scale=0.5]{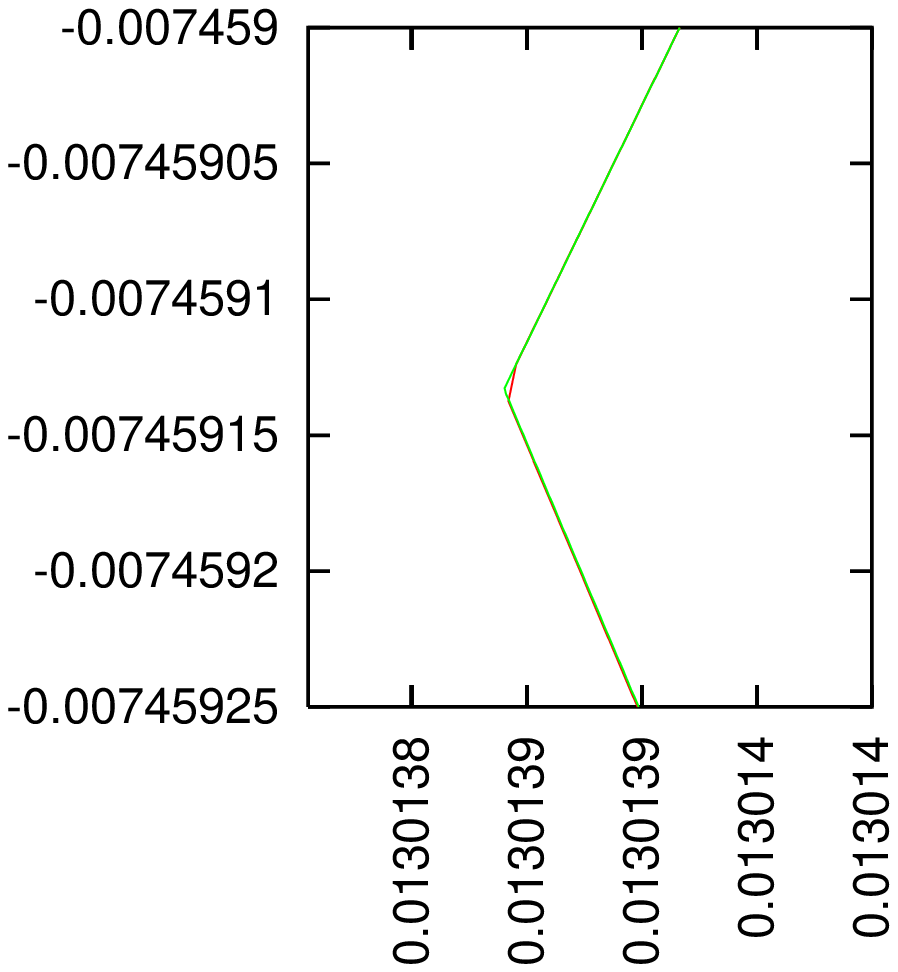}
 ${  }$ \\ [-5cm]
 \hspace{4cm} \includegraphics[scale=0.4]{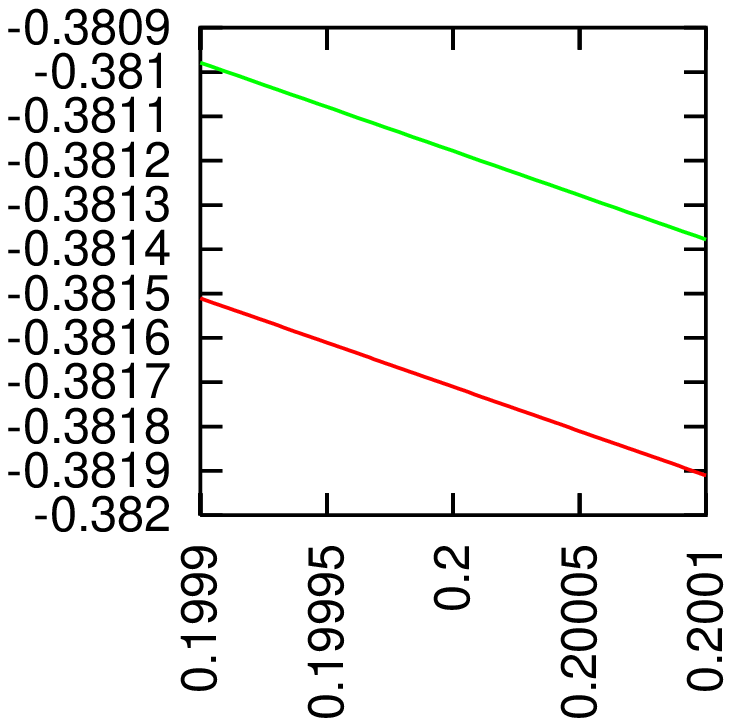}
 ${  }$ \\ [2cm]
\caption{{\bf Left panel:} Paths of Ray 0 (red) and of corrected Backray 0
(green) coincide at the edge of QSS2 to better than $10^{-8}$ in both $X$ and
$Y$. {\bf Right panel:} The same paths midway between QSS2 and the present
observer. The coincidence is only up to $\Delta Y \approx 5 \times 10^{-4}$.}
 \label{lupa}
\end{figure}

The value of $\mu = k^{\vartheta}/k^r$ at $t = t_{\rm fin}$ of (\ref{5.9}), as
calculated by the program that determined Ray 0, was
\begin{equation}\label{d.1}
\mu \df k^{\vartheta}/k^r = 0.0100796385641537794.
\end{equation}
But with this $\mu$ at the initial point, the path of Backray 0 at the entry
point to QSS2 visibly differed (at the scale of the insets in Fig. \ref{uprec})
from that of Ray 0. This had to be caused by numerical errors because there is
no cosmic drift in the Friedmann region \cite{KrBo2011}. Consequently, the $\mu$
in (\ref{d.1}) had to be hand-corrected by trial and error. With $\mu$ given by
(\ref{6.8}), the two rays coincide at the edge of QSS2 up to the level of
accuracy\footnote{See footnote \ref{brline} -- the irregularity of Ray 0 at the
edge of QSS2 is an interpolation by the drawing program.} shown in Fig.
\ref{lupa}.

\section{Calculating the angle $\gamma$ in Fig. \ref{backraylupa}}\label{angamma}

\setcounter{equation}{0}

Let us use $\vartheta$ as the parameter along the arc SF. The unit tangent
vector to SF at $(r, \vartheta) = (r_b, \vartheta_b)$ is
\begin{equation}\label{e.1}
n_{\rm SF}^{\alpha} = \frac 1 {{\cal R}(t_b) \ell}\ \left(0, \left.\dr r
{\vartheta}\right|_{\vartheta = \vartheta_b}, 1, 0\right),
\end{equation}
where ${\cal R}(t_b)$ is given by (\ref{6.15}), and $\left.\dril r
{\vartheta}\right|_{\vartheta = \vartheta_b}$ is found by taking (\ref{6.9}) at
$(r, \vartheta) = (r_b, \vartheta_b)$:
\begin{equation}\label{e.2}
\left.\dr r {\vartheta}\right|_{r = r_b} = - r_b \cot (\vartheta_b + \delta).
\end{equation}
The normalizing factor $\ell$ is
\begin{equation}\label{e.3}
\ell = r_b \sqrt{\frac {\cot^2 (\vartheta_b + \delta)} {1 - k {r_b}^2} + 1}.
\end{equation}
The unit direction vector of Backray 0 at the same point, calculated by analogy
to (\ref{a.3}), is
\begin{equation}\label{e.4}
n^{\alpha}_{\rm ray} = \left(0, \frac {k^r_b} {1 + z_b}, \frac {k^{\vartheta}_b}
{1 + z_b}, 0\right),
\end{equation}
where $z_b$ is given by (\ref{6.14}). Consequently, the angle $\gamma$ between
$n_{\rm SF}^{\alpha}$ and $n^{\alpha}_{\rm ray}$ is given by
\begin{equation}\label{e.5}
\cos \gamma = \frac {r_b {\cal R}(t_b)} {\ell \left(1 + z_b\right)}\ \left[-
\frac {k^r_b \cot (\vartheta_b + \delta)} {1 - k {r_b}^2} + r_b
k_b^{\vartheta}\right].
\end{equation}
The values given in (\ref{6.16}) were calculated from (\ref{e.5}).

\section{Calculating (\ref{7.3}) -- (\ref{7.5})}\label{Downini}

\setcounter{equation}{0}

In the coordinates of the QSS1 region, the line OL obeys the equation $X \equiv
- r \cos \vartheta =$ constant, so the unit tangent vector to it, $n_{\rm
OL}^{\alpha}$, obeys $n_{\rm OL}^r = r \tan \vartheta\ n_{\rm OL}^{\vartheta}$,
$n_{\rm OL}^t = n_{\rm OL}^{\varphi} = 0$ and $g_{\alpha \beta} n_{\rm
OL}^{\alpha} n_{\rm OL}^{\beta} = -1$. The angle $\gamma_N$ in Fig. \ref{sketch}
is between the initial direction of Downray 1 and the unit tangent vector to OL
pointing {\em downward}, so
\begin{eqnarray}\label{f.1}
n_{\rm OL}^r &=& - \frac {\tan \vartheta} {{\cal R}(t) \ell_{\rm OL}}, \qquad
n_{\rm OL}^{\vartheta} = - \frac 1 {r {\cal R}(t) \ell_{\rm OL}}, \nonumber \\
\ell_{\rm OL} &=& \sqrt{\frac {\tan^2 \vartheta} {1 - k r^2} + 1}.
\end{eqnarray}
The unit direction vector of Downray 1 at $P_N$ is, similarly to (\ref{a.3}) and
(\ref{e.4}),
\begin{equation}\label{f.2}
n^{\alpha}_N = \left(0, \frac {k^r_N} {1 + z_N}, \frac {k^{\vartheta}_N} {1 +
z_N}, 0\right).
\end{equation}
At $P_N$, $- g_{\alpha \beta} n_{\rm OL}^{\alpha} n_N^{\beta} = \cos \gamma_N =
\sin \widetilde{\gamma}_N$, so
\begin{equation}\label{f.3}
\frac {\tan \varepsilon} {1 - k {r_1}^2}\ k^r_N + r_1 k^{\vartheta}_N = - \frac
{\ell_N} {{\cal R}_N}\ \left(1 + z_N\right) \sin \widetilde{\gamma}_N,
\end{equation}
where ${\cal D}$ is given by (\ref{7.5}) and
\begin{eqnarray}\label{f.4}
\ell_N &\df& \ell_{\rm OL}(r_1, \varepsilon) = \frac {\cal D} {\sqrt{1 - k
{r_1}^2}}, \nonumber \\
t_N &\df& t_1 + \Delta t_N, \quad {\cal R}_N \df {\cal R}(t_N).
\end{eqnarray}

The vector $k^{\alpha}$ obeys $g_{\alpha \beta} k^{\alpha} k^{\beta} = 0$. Since
Downray 1 prolongs Backray N, Eq. (\ref{3.15}) continues to hold along it.
Consequently, at $(r, \vartheta) = (r_1, \varepsilon)$ we have
\begin{equation}\label{f.5}
\left(1 + z_N\right)^2 - \frac {{{\cal R}_N}^2} {1 - k {r_1}^2}
\left(k^r_N\right)^2 - \left(r_1 {\cal R}_N\right)^2
\left(k^{\vartheta}_N\right)^2 = 0.
\end{equation}
Solving the set (\ref{f.3}), (\ref{f.5}) for $k^r_N$ and $k^{\vartheta}_N$ we
obtain
\begin{eqnarray}
&& k^r_N = \frac {\sqrt {1 - k {r_1}^2} \left(1 + z_N\right)} {{\cal R}_N {\cal
D}} \nonumber \\
&&\ \ \ \ \ \times \left(- \tan \varepsilon \sin \widetilde{\gamma}_N \pm \sqrt
{1 - k {r_1}^2} \cos \widetilde{\gamma}_N\right),\ \ \ \ \ \ \  \label{f.6} \\
&& k^{\vartheta}_N = \frac {1 + z_N} {r_1 {\cal R}_N {\cal D}} \nonumber \\
&& \ \ \ \ \ \times \left(- \sqrt {1 - k {r_1}^2} \sin \widetilde{\gamma}_N
\mp\tan \varepsilon \cos \widetilde{\gamma}_N\right).\ \ \ \ \ \ \ \label{f.7}
\end{eqnarray}
The double sign in (\ref{f.6}) follows from the ambiguity in the definition of
$k^r_N$: its direction is given only by the angle it forms with OL, so it can
point away from QSS1 or into QSS1. We need to follow the second possibility,
which corresponds to the lower sign and implies $k^r_N < 0$. The upper sign
would correspond to $k^r_N > 0$ because $\sqrt {1 - k {r_1}^2} \cos
\widetilde{\gamma}_N \gg \tan \varepsilon \sin \widetilde{\gamma}_N$. This is
because $\sqrt {1 - k {r_1}^2} > 1$, $\cos \widetilde{\gamma}_N \approx 1$ while
$\sin \widetilde{\gamma}_N$ and $\tan \varepsilon$ are both tiny in consequence
of (\ref{7.2}) and (\ref{6.22}).

The situation with the sign of $k^{\vartheta}_N$ is less clearcut. The double
sign in (\ref{f.7}) is coupled with that in (\ref{f.6}), and having chosen
``$-$'' in (\ref{f.6}) we have to choose ``$+$'' in (\ref{f.7}) -- but then
$k^{\vartheta}_N$ might still be positive or negative, depending on the
numerical values of the quantities in it. Since $\pi/2 - \gamma_N =
\widetilde{\gamma}_N > \varepsilon$ as seen from (\ref{6.22}) and (\ref{7.2}),
and $\vartheta$ increases clockwise, the initial direction of Downray 1 is
toward decreasing $\vartheta$, so $k^{\vartheta}_N < 0$ follows. This is
qualitatively reflected in Fig. \ref{sketch}. The Fortran 90 program,using
(\ref{f.7}), found $k^{\vartheta}_N = -3.39090742613403823 \times 10^{-12}$.

Equations (\ref{7.3}) -- (\ref{7.4}) are (\ref{f.6}) -- (\ref{f.7}) with the
signs already chosen.

\section{Numerical calculations of Downrays 1 and 2}\label{inflight}

\setcounter{equation}{0}

Throughout this section, the $r$-coordinate is meant relative to the origin of
QSS1, so $r = r_{\rm S1}$ in all cases.

The difference of the $t$-coordinates of Upray and Downray 1 at the smallest $r$
($\approx 2.23 \times 10^{-22}$) is $\approx 5.437 \times 10^{-6}$. Since
Downray 1 is supposed to represent the ray that reaches the observer 10 min
later than Upray, their time-difference at $r = 0$ should be 10 min $\times (1 +
z)_{\rm 3\ up} = 7.19 \times 10^{-2}$ min = $1.395 \times 10^{-18}$ NTU, where
$(1 + z)_{\rm 3\ up}$ is the upward $1 + z$ between $r = 0$ and the present
observer given by (\ref{5.6}). So clearly, the difference found by the Fortran
program is unrealistically large. If Downray 1 were continued uncorrected, it
would fly above the BB hump and intersect the LSH in the Friedmann region with
$1 + z$ as given in (\ref{7.8}); part of its path is shown in Fig.
\ref{downcor}. Taken literally, this would mean that the frequency on it is no
longer in the gamma range at the observation event. However, this would not
prove the intended point of this paper. It is known that the high-frequency
flash will end if the observer simply waits long enough, but the time-scale of
this process is too long \cite{Kras2016a}. What we wish to show here is that 10
min after the observer saw the gamma flash, the ray coming from the same
direction will be emitted off the symmetry axis of QSS1 and will have a lower
frequency for this reason. Therefore, the $t$-coordinate on Downray 1 at closest
approach to $r = 0$ had to be decreased by hand. The ray on which this was done
is labelled as Downray 2.

\begin{table}[h]
\begin{center}
\caption{Changes in $\Delta \lambda$ in a vicinity of
$r = 0$. \\
 The initial value was $\Delta \lambda = 10^{-12}$. \\
 $r$ in the last line is the closest approach to $r = 0$}
\bigskip
\begin{tabular}{|c|c|}
  \hline \hline
  At $r =$ & $\Delta \lambda$ was multiplied by \\
  \hline \hline
  0.0149 & 100 \\
  \hline
  $2 \times 10^{-6}$ & 1/1000 \\
  \hline
  $10^{-10}$ & 1/1000 \\
  \hline
  $2 \times 10^{-13}$ & 1/100 \\
  \hline
  $2 \times 10^{-15}$ & 1/100 \\
  \hline
  $5 \times 10^{-17}$ & 1/100 \\
  \hline
  $2 \times 10^{-19}$ & 1/100 \\
  \hline
  $ \approx 2.23 \times 10^{-22}$ & 1/1000 \\
  \hline \hline
\end{tabular} \\
 \label{reductions}
\end{center}
\end{table}

\begin{figure}[h]
 \includegraphics[scale=0.6]{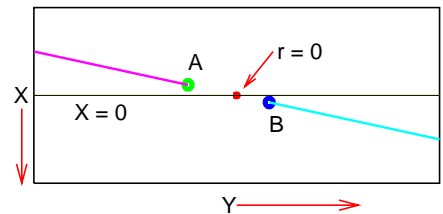}
\caption{When Downray 2 passes near $r = 0$, at the point A of coordinates
$(r_A, \vartheta_A)$ the numerical program jumps to $(r_B, \vartheta_B)$ with
$r_B < 0$. This is interpreted as the jump to point B of coordinates $(r,
\vartheta) = (-r_B, \vartheta_A + \pi)$. This graph is a sketch; for the real
situation see Fig. \ref{atcent}.}
 \label{jump}
\end{figure}

\begin{figure}[h]
 \includegraphics[scale=0.5]{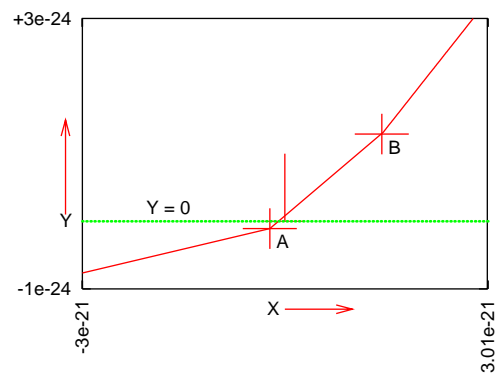}
\caption{The situation from Fig. \ref{jump} in real scale. The segment AB is an
interpolation by the plotting program. The $(X, Y)$ coordinates of points A and
B are $(-2.23 \times 10^{-22}, -1.073 \times 10^{-25})$ and $(1.4367 \times
10^{-21}, 1.29 \times 10^{-24})$, respectively. The longer vertical stroke marks
$X = 0$. }
 \label{atcent}
\end{figure}

Downray 2 runs near to the symmetry axis of QSS1 and is nearly parallel to it.
When the ray approaches $r = 0$, at a certain step the next $r$ becomes
negative. In an attempt to prevent this, the step $\Delta \lambda$ in the affine
parameter $\lambda$ was gradually decreased on approaching the origin. Table
\ref{reductions} shows the reductions in $\Delta \lambda$ and the values of $r$
at which they were done. But the jump to $r < 0$ persisted. So, when the Fortran
program reached the limit of its precision, the jump to $r < 0$ was interpreted
as the jump from the last $(r, \vartheta)$ where $r$ was still positive to $(r,
\pi + \vartheta)$ (see Figs. \ref{jump} and \ref{atcent}), and the program
continued from there. As Downray 2 was receding from the origin and passed
through the values of $r$ from Table \ref{reductions} in a reverse order,
$\Delta \lambda$ was multiplied by the reverse of the factor from the table at
each border value of $r$, except the one in the last line. In addition, at $r =
0.01$ and $X > 0$, $\Delta \lambda$ was multiplied by 100. Figures \ref{downcor}
-- \ref{microray2} show the resulting graphs.

In order to locate the intersection of Downray 2 with the LSH with a
satisfactory precision, more modifications of $\Delta \lambda$ were made. It was
multiplied by 1/100 at step 114,685,150, and then again by 1/1000 at step
115,226,000. The LSH and BB were reached in steps 114,698,353 and 115,634,944,
respectively, but the last number is not reliable. Since the ray becomes tangent
to a surface of constant $r$ when approaching the BB, it is difficult to locate
the intersection point -- the result strongly depends on the value of $\Delta
\lambda$. But the exact point where the ray hits the BB has no physical meaning
because the QSS metric does not apply at times before the LSH.

\section{Details for Sec. \ref{nearby}}\label{apnearby}

\setcounter{equation}{0}

Table \ref{zondownrays} gives the redshifts/blueshifts on Downrays 5 to 13
between the LSH and the point $P_N$ ($1 + z_P$), and between the LSH and the
present observer ($1 + z_o$).

\begin{table}[h]
\begin{center}
\caption{Redshift from the LSH on Downrays 5 to 13}
\bigskip
\begin{tabular}{|c|c|c|}
  \hline \hline
  ray & $1 + z_P$ & $1 + z_o$ \\
  \hline \hline
  5 & 0.070480436257171 & 5.91956865287204 \\
  \hline
  6 & 0.110226238143066 & 9.25777164112425 \\
  \hline
  7 & 0.024772171188802 & 2.08058541944529 \\
  \hline
  8 & 0.113160537766071 & 9.50422000309368 \\
  \hline
  9 & 0.104598722390451 & 8.7851232352436 \\
  \hline
  10 & 0.020657768217553 & 1.73502157013793 \\
  \hline
  11 & 0.011190786419531 & 0.939900942842124 \\
  \hline
  12 & 0.330430215246075 & 27.7524437702868 \\
  \hline
  13 & 0.012570655453927 & 1.05579451437558 \\
  \hline \hline
\end{tabular} \\
 \label{zondownrays}
\end{center}
\end{table}

\begin{figure}[h]
 \hspace{-5mm} \includegraphics[scale=0.5]{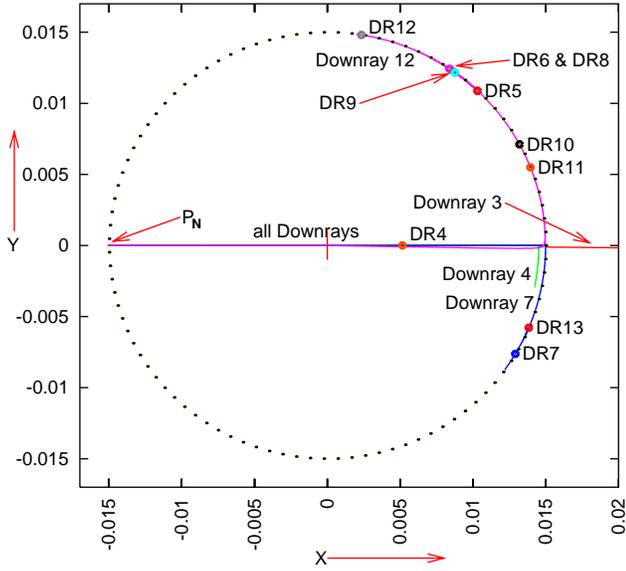}
\caption{The intersections of Downrays 3 to 13 with the LSH projected on the
$(X, Y)$ coordinate plane (marked with large dots). Projections of selected rays
are shown in full; the remaining ones follow parts of the paths of Downray 7 or
Downray 12. Downray 3 intersects the LSH in the Friedmann region beyond the
right margin of the figure. }
 \label{downrecbis}
\end{figure}

Figure \ref{downrecbis} shows where Downrays 3 to 13 intersect the LSH; the
intersection points are marked with large dots. Four typical examples of the
rays are displayed in full. Downray 3 escapes from the QSS1 region and
intersects the LSH in the Friedmann region, beyond Fig. \ref{downrecbis}.
Downray 4 intersects the LSH between $r = 0$ and the boundary of the QSS1
region, and hits the BB still inside QSS1. Downray 7 turns toward $Y < 0$ near
the boundary of the QSS1 region, and intersects the LSH near the steep slope of
the BB hump. Downray 12 turns toward $Y > 0$, and intersects the LSH near
another point of the steep slope of the BB hump. Downray 13 follows a part of
the path of Downray 7, the other rays follow parts of the path of Downray 12.
They intersect the LSH at various $\vartheta$, but at $r$ nearly on the boundary
of the BB hump.

\begin{figure}[h]
 \hspace{-1cm} \includegraphics[scale=0.5]{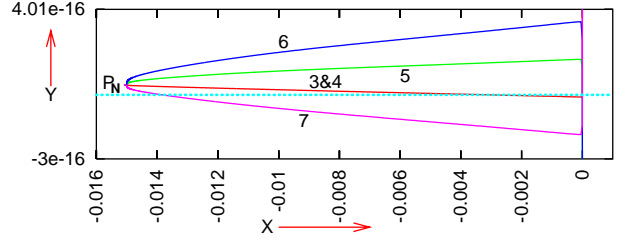}
\caption{Downrays 3 to 7 in the segment $X \in (- r_1, 0)$. Near to $r = 0$ they
all begin to aim at $r = 0$. The passage through $r = 0$ is handled as described
in Appendix \ref{inflight}. The dotted horizontal line marks $Y = 0$.}
 \label{dr3to7}
\end{figure}

\begin{figure}[h]
 \hspace{-1cm} \includegraphics[scale=0.5]{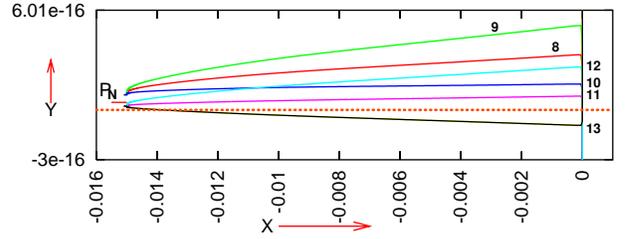}
\caption{Downrays 8 to 13 in the segment $X \in (- r_1, 0)$. Near to $r = 0$
they behave just as those in Fig. \ref{dr3to7}. The short horizontal dash marks
the $Y$ coordinate of the point $P_N$.}
 \label{dr8to13}
\end{figure}

Figures \ref{dr3to7} and \ref{dr8to13} show the rays between their initial
points and the neighbourhood of $r = 0$. In this segment, they stay within the
strip $-2 \times 10^{-16} < Y < 6 \times 10^{-16}$ and pass through $r = 0$ in
ways similar to that shown in Figs. \ref{jump} and \ref{atcent}.

\begin{figure}[h]
 \includegraphics[scale=0.5]{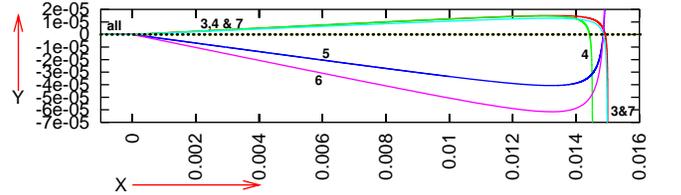}
\caption{Downrays 3 to 7 in the segment $X \in (- 0.001, 0.016)$. At the scale
of this figure, all rays coincide for $X < 0$. The dotted horizontal line marks
$Y = 0$.}
 \label{dr3to7ri}
\end{figure}

\begin{figure}[h]
 \includegraphics[scale=0.5]{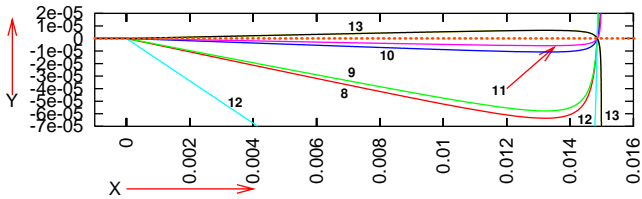}
\caption{Downrays 8 to 13 in the segment $X \in (- r_1, 0)$. The minimum $Y$ on
Downray 12 is $\approx -1 \times 10^{-4}$. }
 \label{dr8to13ri}
\end{figure}

Figures \ref{dr3to7ri} and \ref{dr8to13ri} show the rays between $r = 0$ and $r
= r_F$ given by (\ref{2.23}). Here they depart from the $Y = 0$ line much
further than in the $X < 0$ segment: most stay in the strip $-7 \times 10^{-5} <
Y < 2 \times 10^{-5}$, except for Downray 12 that at the farthest distance from
the $Y = 0$ axis has $Y \approx -1 \times 10^{-4}$. Close to the boundary of
QSS1 they bend around and become nearly tangent to surfaces of constant $r$.
Downray 3 is an exception: it follows the nearly-constant-$r$ path for a short
while, then escapes to larger $r$, as seen in Fig. \ref{downrecbis}. Downray 4
hits the BB at $r$ distinctly smaller than the $r_F$ of (\ref{2.23}), the
remaining rays hit it at $r$ very close to $r_F$.

{\bf Acknowledgement} For some calculations, the computer algebra system
Ortocartan \cite{Kras2001,KrPe2000} was used.

\end{document}